\documentclass[preprint,authoryear,12pt]{elsarticle}

\usepackage{graphics}
\usepackage{multirow}
\usepackage{amssymb}
\usepackage{lineno}
\usepackage{lscape}
\usepackage{color}

\journal{Icarus}

\begin{document}

\begin{frontmatter}

\title{A population of main belt asteroids co-orbiting with Ceres and Vesta}

\author[addr1]{Apostolos A.~Christou\corref{cor1}}
\ead{aac@arm.ac.uk}
\ead{Fax: +44 2837 522928}
\address[addr1]{Armagh Observatory, College Hill,
           Armagh BT61 9DG, Northern Ireland, UK}
\cortext[cor1]{Corresponding author}
\author[addr2]{Paul Wiegert}
\ead{pwiegert@uwo.ca}
\address[addr2]{The University of Western Ontario, 
Department of Physics and Astronomy, London, Ontario N6A 3K7, Canada}

\begin{abstract}
We have carried out a search for Main Belt Asteroids (MBAs) co-orbiting with the large MBA Vesta and the dwarf planet Ceres. 
Through improving the search criteria used in \citet{Christou2000b} and numerical integrations of candidate coorbitals, 
we have identified approximately 51 (44) objects currently in co-orbital libration with Ceres (Vesta). 
We show that these form part of a larger population of transient coorbitals; 129 (94) MBAs undergo episodes of 
co-orbital libration with Ceres (Vesta) within a 2 Myr interval centred on the present. The lifetime in the 
resonance is typically a few times $\sim 10^{5}$ yr but can exceed $2 \times 10^{6}$ yr.
The variational properties of the orbits of several co-orbitals were examined. It was found that their 
present states with respect to the secondary are well determined but knowledge of it is lost typically after 
$\sim 2 \times 10^{5}$ years. Objects initially deeper into the coorbital region maintain their coorbital state for longer. 
Using the model of \citet{Namouni.et.al1999} we show that their dynamics are similar to those of temporary 
coorbital NEAs of the Earth and Venus. As in that case, the lifetime of resonant libration is dictated by 
planetary secular perturbations, the inherent chaoticity of the orbits and close encounters with massive objects 
other than the secondary. In particular we present evidence that, while in the coorbital state, close encounters 
with the secondary are generally avoided and that Ceres affects the stability of tadpole librators of Vesta. 
Finally we demonstrate the existence of Quasi-satellite orbiters of both Ceres and Vesta and conclude that 
decametre-sized objects detected in the vicinity of Vesta by the DAWN mission may, in fact, belong to 
this dynamical class rather than be bona-fide (i.e.~keplerian) satellites of Vesta.      
\end{abstract}
   

\end{frontmatter}

\linenumbers

\section{Introduction}
The coorbital resonance, where the gravitational interaction between two 
bodies with nearly the same orbital energy leads to stable and predictable 
motion, is ubiquitous in the solar system. Objects attended by known 
co-orbital companions include Jupiter, Mars as well as the saturnian 
satellites Tethys, Dione, Janus and Epimetheus \cite[see][for a review]{Christou2000a}. 
More recently, the planet Neptune was added to this list 
\citep{SheppardTrujillo2006} while an additional coorbital of Dione 
was discovered by the {\it Cassini} mission \citep{Murray.et.al2005}.
In all these cases, the motion has been shown to be stable against all but the most 
slow-acting perturbations \citep{Lissauer.et.al1985,Levison.et.al1997,Brasser.et.al2004c,Scholl.et.al2005}.

The discovery of a coorbital attendant of the Earth on a highly inclined and eccentric
orbit \citep{Wiegert.et.al1997,Wiegert.et.al1998} motivated new theoretical work in the field.
\citet{Namouni1999}, using Hill's approximation to the Restricted Three Body Problem (R3BP) 
showed analytically that the introduction of large eccentricity and inclination modifies 
considerably the topology of coorbital dynamics near the secondary mass. It results in
the appearance of bounded eccentric orbits \cite[Quasi-Satellites;][]{MikkolaInnanen1997} 
and, in three dimensions, ``compound'' orbits and stable transitions between the different modes 
of libration. Further, he demonstrated numerically that these results hold when the full R3BP 
is considered. In this case, the appearance of new types of compound orbits, such as asymmetric modes or
compounds of tadpoles and retrograde satellites, was shown in \citet{Namouni.et.al1999} 
to be due to the secular evolution of the coorbital potential. Such types of coorbital 
libration were identified in the motion of the object 
highlighted by \citet{Wiegert.et.al1997} as well as other near-Earth asteroids but the 
secular forcing of the potential in that case is provided by planetary secular perturbations 
\citep{Namouni.et.al1999,Christou2000a}. The expected characteristics of the population
of co-orbitals of Earth and Venus were investigated by \citet{MoraisMorbidelli2002} and
\citet{MoraisMorbidelli2006} respectively.

\citet{Christou2000b}, motivated by Pluto's ability, as demonstrated by 
\citet{YuTremaine1999} and \citet{Nesvorny.et.al2000}, to 
trap other Edgeworth-Kuiper Belt objects in co-orbital motion with itself,
demonstrated in turn that main belt asteroids can co-orbit with the dwarf planet 1 Ceres and the large
Main Belt Asteroid (MBA) 4 Vesta.
Four such asteroids were identified, two co-orbiting with Ceres and two with Vesta. 

Here we report the results of a search for additional coorbitals of these two massive asteroids.
This was motivated partly by the large growth in the number of sufficiently well-known MBA orbits
during the intervening decade, but also a refinement of the search criterion used in the work by Christou.  
As a result, we find over 200 new transient co-orbital MBAs of Ceres and Vesta. In this work we examine 
their ensemble properties, use existing dynamical models to understand how they arise and identify similarities
with co-orbital populations elsewhere in the solar system, in particular the transient coorbital NEAs of Earth and Venus.

The paper is organised as follows:
In the next Section we expose our Search methodology, in particular those aspects which differ
from the search carried out by \citet{Christou2000b}. In Section 3 we describe the statistics
of coorbital lifetime and orbital element distribution found in our integrations.  In addition, we examine the
robustness of the dynamical structures we observe. 
In Section 4 we investigate the degree to which the model of \citet{Namouni.et.al1999} 
can reproduce the observed dynamics. Section 5 deals with the effects of additional massive asteroids
in the three-body dynamics while Section 6 focuses on the stability of so-called Quasi-Satellite orbits.
Finally, Section 7 summarises our conclusions and identifies further avenues of investigation. 
  
\section{Search Methodology}
\begin{figure}
\centering
\includegraphics[width=11cm,angle=0]{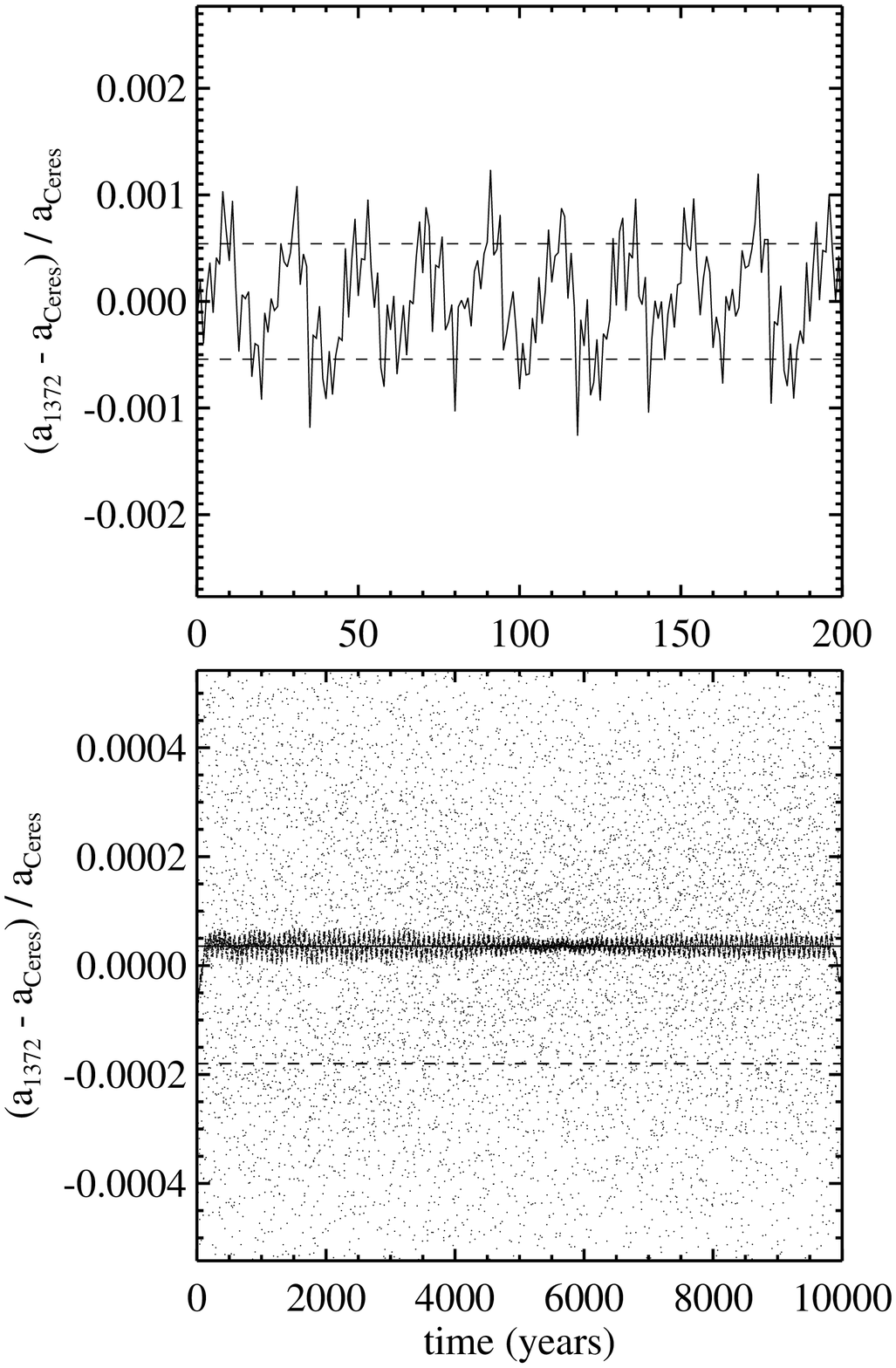}
\caption[Comparison of proper, averaged and osculating semimajor axis for asteroid 1372 Haremari during a numerical $10^{4}$ yr numerical integration of its orbit. See text for details.]{Christou and Wiegert 2010, 
Coorbitals of Ceres and Vesta}
\label{fig:prpr_vs_osc}
\end{figure}

\citet{Christou2000b} searched for candidate Ceres coorbitals by employing the osculating semimajor 
axis $a_{r}$ of the asteroid relative to that of Ceres (equal to $\left(a - a_{\rm Ceres}\right) / a_{\rm Ceres}$) 
to highlight objects that merited further investigation. This method, although it led to the successful 
identification of two coorbitals, 1372 Haremari and 8877 Rentaro, ignores the existence of high frequency 
variations in $a$ due to perturbations by the planets, especially Jupiter. This is illustrated in the 
upper panel of Fig.~\ref{fig:prpr_vs_osc} where the evolution of $a_{r}$ for asteroid 1372 Haremari, 
one of the objects identified by Christou, is depicted. A periodic term of amplitude $0.001$ apparently causes 
$a_{r}$ to move in and out of Ceres' coorbital zone (bounded by the dashed horizontal lines) every 20 years. 
On the other hand, an object with $a_{r}\leq \epsilon$ would have a conjunction with Ceres every 
$\geq 6 \times 10^{4}$ years. Here $\epsilon = {\left(\mu/3\right)}^{1/3}$ 
denotes Hill's parameter in the Circular Restricted Three-Body Problem for a secondary body with a mass
$\mu$ scaled to that of the central body (in this case, the Sun). 
This suggests that coorbital motion should be insensitive to these variations
and that a low-pass filtered semimajor axis would work better as an indicator of libration in the 1:1 resonance 
with Ceres. We have chosen this to be the synthetic proper semimajor axis (hereafter referred to as 
``proper semimajor axis'') as defined by \citet{KnezevicMilani2000}. Synthetic proper elements are 
numerically-derived constants of an asteroid's motion under planetary perturbations. The proper semimajor axis, 
in particular, is invariant with respect to secular perturbations up to the second order in the masses 
under Poisson's theorem and, hence, an appropriate metric to use in this work. In the bottom panel of Fig.~\ref{fig:prpr_vs_osc}, we plotted the 
proper relative semimajor axis $a_{r,\mbox{ }p}$ $=\left( a_{p} - a_{p\mbox{, }{\rm Ceres}}\right)/a_{p\mbox{, }{\rm Ceres}}$; bold horizontal line) superimposed on the 
$a_{r}$ history of the asteroid but filtered with a 256-point running-box average. 
The slopes at the beginning and the end of this time series are the result of incomplete averaging.
The dots in the bottom panel represent the unaveraged osculating semimajor axis  sampled with a constant timestep of 1 yr. 
It is observed that the proper semimajor axis agrees with the average of its osculating counterpart, 
whereas the osculating value sampled at $t=0$ ( Julian Date 2451545.0; dashed horizontal line) does not.   
 
For our search we used the database of synthetic proper elements for 185546 numbered asteroids 
computed by Novakovi\'{c}, Kne\v{z}evi\'{c} and Milani that was available as of 09/2008 at the 
{\it AstDys} online information service (http://hamilton.dm.unipi.it/$\sim$astdys/propsynth/numb.syn). 
A total of 648 and 514 main-belt asteroids respectively with proper semimajor axes within 
$\pm a \epsilon$ of those of Ceres and Vesta were identified. Table ~\ref{tab:massive} shows 
the parameters relevant to this search for the massive asteroids. The second row provides the mass ratio $\mu$ 
of the asteroid relative to the mass of the Sun, the third row its Hill parameter $\epsilon$, the fourth row the 
asteroids' synthetic proper semimaxor axis $a_{p}$ as given in the database and the fifth row the product 
$a_{p} \times \epsilon$ which is the half-width of the coorbital region in AU.

The state vectors of these asteroids at Julian Date 2451545.0 were retrieved from
HORIZONS \citep{Giorgini.et.al1996} and numerically integrated one million years in the past and 
in the future using a model of the solar system consisting of the 8 major planets and the asteroids 
1 Ceres, 4 Vesta, 2 Pallas and 10 Hygiea. Mass values adopted for these four asteroids were taken from 
\citet{Konopliv.et.al2006}. Their initial state vectors as well as planetary initial conditions and 
constants were also retrieved from HORIZONS. The integrations were carried out using the ``hybrid'' scheme 
which is part of the MERCURY package \citep{Chambers1999}.  This scheme is based on a second-order mixed 
variable symplectic (MVS) algorithm; it switches to a Bulirsch-Stoer scheme within a certain distance from a 
massive object. For all the integrations reported here, this distance was 5 Hill radii.  A time step of 4 days or
$\sim 1/20$th of the orbital period of Mercury was chosen in order to mitigate the effects of stepsize resonances \citep{WisdomHolman1992}. 
Trials of this scheme vs  an RA15 RADAU integrator \citep{Everhart1985} included in the same package 
 and with a tolerance parameter of $10^{-13}$ showed the results to be indistinguishable from each other 
while the hybrid scheme 
was significantly faster. 

\section{Results}
\subsection{Population Statistics}
In our runs, we observed a total of 129 and 94 asteroids enter in one of the known libration modes of 
the 1:1 resonance with Ceres or Vesta respectively at some point during the integrations. 
A full list of these asteroids is available from the corresponding author upon request.
Table \ref{tab:statistics} shows a statistical breakdown of the observed population according to 
different types of behaviour. The first row identifies the secondary (Ceres or Vesta). 
The top part of the Table shows the number of asteroids that were captured into co-orbital libration 
at some point during the $2 \times 10^{6}$ yr simulation (second column), the number of asteroids 
that were found to be co-orbiting at present ( ``current'' coorbitals; third column) and the number of current 
$\mbox{L}_{4}$ and $\mbox{L}_{5}$ tadpole librators (fourth and fifth rows respectively). 
The bottom part of the table shows the number of current horseshoe librators (second column), 
the number of objects currently in transition between two distinct libration modes (third column) 
and the number of current Quasi-Satellite (QS) librators. The fifth column shows the 
number of asteroids that were librating for the full duration of the numerical integration, 
although not necessarily in the same mode (``persistent'' coorbitals). 
Numbers in brackets refer to MBAs that remained in the same libration mode throughout the integration 
(``single mode persistent'' coorbitals). 
A total of 95 MBAs are currently in the 1:1 resonance with either Ceres or Vesta with the ratio of objects 
corresponding to the two asteroids (51/44) being roughly the same as for the original sample (129/94). 
The number of MBAs currently in tadpole orbits is 19 and 12 for the two asteroids respectively 
while horseshoes are evenly matched at 20 and 21. 11 objects for each asteroid are in the process of 
transitioning between different modes of libration. One object, MBA 76146, is currently in a Quasi-Satellite 
(QS) orbit around Ceres with a guiding centre amplitude of $\sim 10^{\rm o}$ in $\lambda_{r}$ (rightmost 
panels in Fig~\ref{fig:ceres_expls}) while two other MBAs, 138245 and 166529, are currently transitioning 
into such a state.

\subsection{Some Examples}
\begin{figure}
\centering
\includegraphics[width=10cm,angle=90]{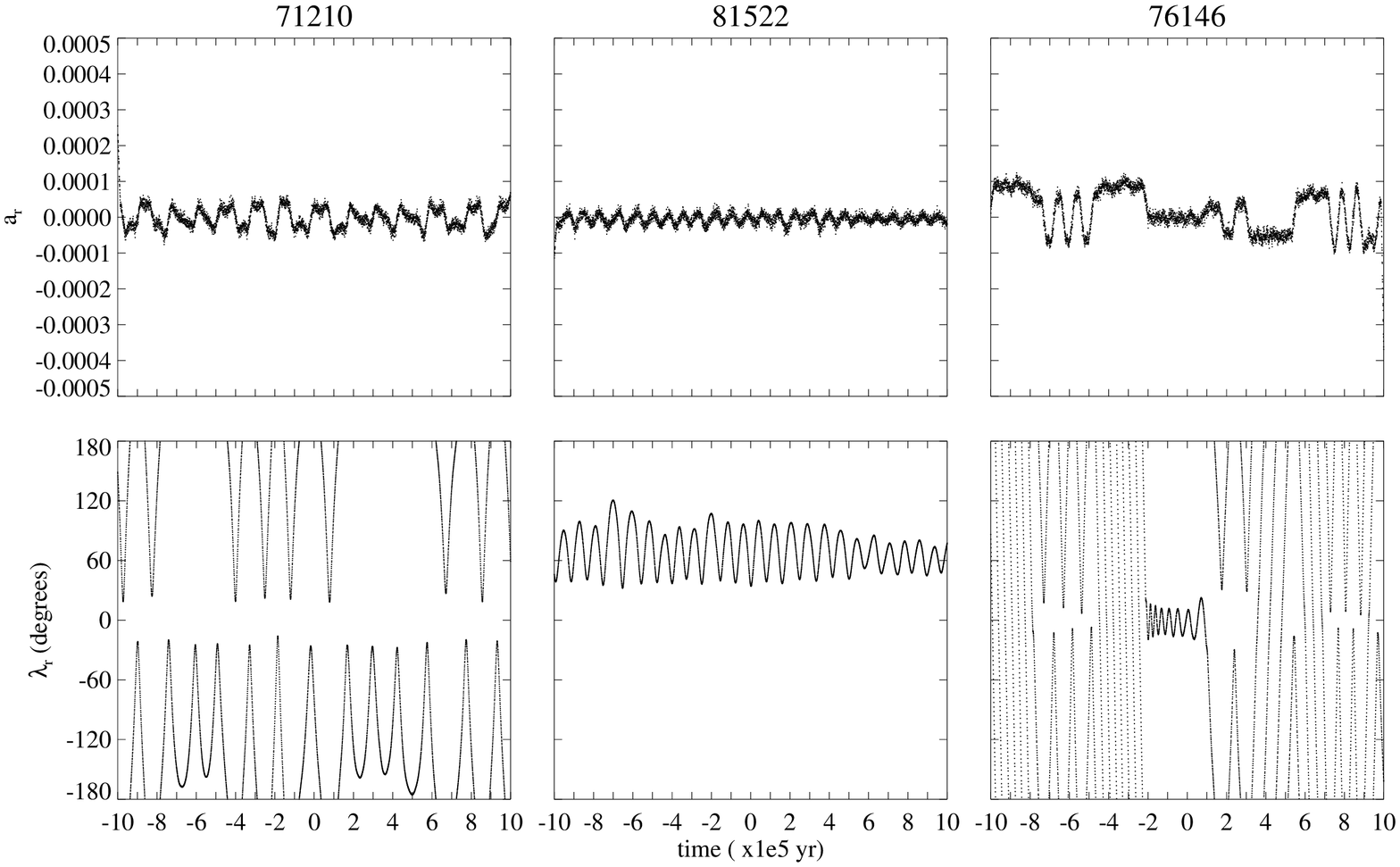}
\caption[Examples of dynamical evolution of specific asteroids coorbiting with Ceres as determined 
by our numerical integrations over $2 \times 10^{6}$ yr. The upper row shows the evolution of the relative 
semimajor axis $a_{r}$ while the bottom row that of the relative longitude $\lambda_{r}$. 
]{Christou and Wiegert 2010, Coorbitals of Ceres and Vesta}
\label{fig:ceres_expls}
\end{figure}

In most cases, the librations were transient i.e. changing to another mode or to circulation of the critical argument 
$\lambda_{r} = \lambda - \lambda_{\rm Ceres/Vesta}$. In that sense, there is a similarity to what is observed 
for co-orbitals of the Earth and Venus \citep{Namouni.et.al1999,Christou2000a,MoraisMorbidelli2002,MoraisMorbidelli2006}. 
 Compound modes (unions of Quasi-Satellite and Horseshoe or Tadpole librations) also appear, although they are rare.
Examples of different types of behaviour are shown in Figs~\ref{fig:ceres_expls} and \ref{fig:vesta_expls} 
for Ceres and Vesta respectively. 

In the former case, asteroid 71210 (left column) is currently ($t=0$ at Julian Date 2451545.0) 
in a horseshoe configuration with Ceres, transitioning to tadpole libration at times. The position of Ceres at 
$\lambda_{r}=0$ is avoided. Asteroid 81522 (centre column) librates around the $\mbox{L}_{4}$ triangular equilibrium 
point of Ceres for the duration of the integration. In the right column we observe stable transitions of the 
orbit of asteroid 76146 between different libration modes. The maximum excursion of the guiding centre of 
the asteroid from that of Ceres during the integration is $a_{r}=10^{-4}$ or $0.2\epsilon$, well within 
the coorbital region. This asteroid becomes a Quasi-Satellite of Ceres at $t=-2 \times 10^{5}$ yr and 
transitions into a horseshoe mode at $t=+10^{5}$ yr.

In the case of Vesta, asteroid 22668 (left column) transitions from a passing to  a horseshoe mode and back again. 
It is currently a horseshoe of Vesta. Near the end of the integration it enters into Quasi-Satellite 
libration with Vesta, where it remains.  At $t=+5 \times 10^{5}$ yr it executes half a libration in a 
compound Horseshoe/Quasi Satellite mode. Asteroid 98231 (centre column) is an $\mbox{L}_{5}$ tadpole of Vesta 
at $t=-1 \times 10^{6}$ yr. The libration amplitude increases gradually until transition into 
horseshoe libration occurs at $t=-3 \times 10^{5}$ yr. During this period the libration 
amplitude begins to decrease until the reverse transition back into $\mbox{L}_{4}$ libration takes place at 
$t=+6 \times 10^{5}$ yr. This behaviour should be compared with the case of asteroid 71210 in Fig.~\ref{fig:ceres_expls}. 
The evolution of asteroid 156810 (right column) is similar to that of 98231 except that the rate of 
increase of the libration amplitude reverses sign before transition into horseshoe libration takes 
place. As a result, the asteroid librates around the $\mbox{L}_{4}$ equilibrium point of Vesta for the duration 
of the integration.

\subsection{Variational Properties}
Although these results can be regarded statistically, their value is increased if we establish their 
robustness against the ephemeris uncertainties of the asteroids. To investigate this, we have picked 
6 asteroids in the sample, populated their 1-sigma uncerainty ellipsoids with 100 clones per object 
using states and state covariance matrices retrieved from the {\it AstDys} online orbital information service, 
and propagated them forward in time for $10^{6}$ yr under the same model of the solar system as before but 
using  an integrator developed by one of us (PW) with a 4-day time step.  This code utilises 
the same algorithm as the HYBRID code within MERCURY. In the original integrations, 
65313 and 129109 persist as $\mbox{L}_{5}$ Trojans of Ceres, 81522 and 185105 as $\mbox{L}_{4}$ Trojans of 
Ceres and 156810 persists as an $\mbox{L}_{5}$ Trojan of Vesta. Asteroid 76146 is a Quasi-Satellite of Ceres until 
+${10}^{5}$ yr. Their Lyapounov times, as given by the proper element database of Novakovi\'{c} et al, are: $3 \times 10^{6}$, $5 \times 10^{5}$,  
$2 \times 10^{6}$,  $6 \times 10^{5}$,  $10^{5}$ and $5 \times 10^{5}$ yr respectively.

The results of this exercise are shown in Fig.~\ref{fig:stability}. We find that all clones of 65313 and 
81522 remain as Trojans of Ceres for the full integration. 76146 persists as a Quasi-Satellite of Ceres 
until $\sim +6 \times 10^{5}$ yr.  Three clones of 129109 escape from libration 
around $\mbox{L}_{5}$ after $\sim 5 \times 10^{5}$ yr, but most remain in libration until the end. 
Those of 185105 begin to diverge at $\sim + 3 \times 10^{5}$ yr until knowledge of the object's state 
is lost near the middle of the integration timespan. The clones of the Vesta Trojan 156810 suffer a 
similar fate, but divergence is more general in nature; most of the clones enter into other libration 
modes by $ +6 \times 10^{5}$ yr. We conclude that the present state of these objects, as determined by 
the original integrations, is robust. The differences we observe in their evolution could be due to 
several causes. Firstly, although we use the same force model and the same type of integrator, 
the initial states of the asteroids in the new integrations were taken from AstDys, not HORIZONS, 
and correspond to a later epoch of osculation. 
In addition, the volumes of space sampled by the object's ephemeris uncertainties, being proportional to 
the eigenvalue product of the respective covariance matrices, are smallest for 65313 and 81522. 
We therefore expect that, assuming similar Lyapounov times, clones of low-numbered asteroids - generally 
those with longer observational arcs - would be slower to disperse than those of high-numbered ones. 
Interestingly, the clones disperse faster than one would expect from their Lyapounov times, however 
some differences would be expected since the model used to compute these did not include the gravitational 
attraction of large asteroids. In Section 5, the role of impulsive perturbations during close encounters 
is investigated in detail. 

\subsection{Proper element distribution}
In order to understand the ensemble properties of the population and how these might differ from those of 
other MBAs, we have compared the distribution of their proper elements to that of the broader population. 
Fig~\ref{fig:histo_ap} shows the distribution of the relative proper semimajor axis (blue curve)
$a_{r\mbox{, }p}=\left(a_{p} - a_{p\mbox{, }{\rm Ceres/Vesta}}\right)/a_{p \mbox{, }{\rm Ceres/Vesta}}$ of these MBAs, 
scaled to $\epsilon$, superposed on that of all MBAs in the interval $[-2 \epsilon\mbox{, }+2\epsilon]$ (red curve). 
The width of each bin for both panels is 0.06. The sharp peak at $a_{r}=-1.2$ in the plot for Vesta 
reflects an increased concentration of asteroid proper elements probably associated with the 36:11 mean 
motion resonance with Jupiter. Although interesting in its own right, 
we do not, at present, have reason to believe that its existence near the co-orbital region of Vesta is 
anything more than coincidental. Hence, we refrain from dicussing it further in this paper.

In both cases, the distributions appear to be centred at $a_{r\mbox{, }p}=0$. Gaussian fits to the centre $\mu$ 
and the standard deviation $\sigma$ of the distribution give a Full Width at Half Maximum 
(FWHM; $2 \sqrt{2 \log 2} \sigma$) of $0.334 \pm 0.033$ and a centre at  $-0.053 \pm 0.017$ for the 
Ceres distribution. The slight offset to the left is probably due to slightly higher counts for the bins 
left of $a_{r\mbox{, }p}=0$. The Vesta distribution is slightly narrower (FWHM of $0.292 \pm  0.028$) but more 
symmetric around the origin (centre at $ -0.006 \pm 0.014$). No cases of coorbital libration were observed 
for asteroids with $|a_{r\mbox{, }p}|>0.42$ while only two cases (both with Vesta) had  $|a_{r\mbox{, }p}|>0.30$. 

 Seeking additional insight into the dependence of the coorbital state on the semimajor axis, we examined 
the distributions of different types of co-orbitals - as observed in our simulations - normalised 
to the total number of MBAs in each bin. In Fig.~\ref{fig:histo_ap2} we show the distributions 
of  all coorbitals (red curve), current coorbitals (blue curve) and persistent 
coorbitals (gray curve). 
Fitted values of the Gaussian parameters $\left(\mu,\mbox{ FWHM} \right)$ for the three populations 
are given in Table~\ref{tab:mufwhm}. The distributions for Vesta coorbitals are consistently narrower than those of Ceres 
implying that this is a real difference between the two. 

Interestingly, both cases exhibit a 
hierarchy in the three populations: the persistent population is embedded into the current one, which in turn 
is embedded into the distribution of  all objects. One important consequence of this observation 
is that one can robustly define the boundaries of each population. This is, of course, partly due to the 
criteria used to define each population but the fact that there are clear differences between the three 
populations (ie no two populations coincide) is not a trivial one. Hence, persistent coorbitals are confined 
in the domain $|a_{r\mbox{, } p}|<0.12$, current coorbitals within $|a_{r\mbox{, }p}|<0.24$ and  all coorbitals 
in the domain $|a_{r\mbox{, } p}|<0.42$. The shape of the distributions in Figs.~\ref{fig:histo_ap} and \ref{fig:histo_ap2} 
can be partly attributed to the coorbital dynamics (see Section 4). 
However, they must also be affected by chaotic `noise' in the determination of the proper elements which we 
used in our search \citep{MilaniKnezevic1994}. In their 2 Myr integrations, \citet{KnezevicMilani2000} 
regarded the derived proper elements of MBAs with $\sigma_{a_{p}}<3 \times 10^{-4}$, $\sigma_{e_{p}}<3 \times 10^{-3}$ 
and $\sigma_{I_{p}} < 10^{-3}$ as `good'. All but four  of the asteroids considered here belong to this catelogy. 
The bound for the proper semimajor axis corresponds to 20\% of the width of the coorbital region of Ceres and 35\% of 
that of Vesta (Table~\ref{tab:massive}). It is also comparable to the fitted widths of the distributions of the coorbitals 
found here. Hence, the actual distributions are likely significantly altered by a convolution with an error function. 
On the other hand, this convolution does not completely smear out the true distribution of $a_{r\mbox{, }p}$ 
since, in that case, the observed sorting of the populations according to residence time in the 
resonance would not occur (Fig.~\ref{fig:histo_ap2}). 

The distribution of the proper eccentricities and inclinations of individual coorbitals in relation to 
those of other MBAs are shown in Fig.~\ref{fig:aei}. 
 Plus symbols denote MBAs that have tested negative for co-orbital motion within the period 
$\mbox{[}-10^{6}\mbox{ yr, }10^{6}\mbox{ yr]}$. Asterisks and squares refer to the respective populations of 
current and persistent coorbitals while the filled circle marks the location of the secondary (either Ceres or Vesta).
In the interests of clarity, we  have not plotted the distribution of all coorbitals. Instead, we show as triangles 
those persistent coorbitals that remained in either $\mbox{L}_{4}$ or $\mbox{L}_{5}$ tadpole libration for the full simulation. 
Their location deep into the coorbital region are in agreement with the theoretical upper 
limit - $\sqrt{(8/3) \mu_{\rm Ceres/Vesta}}$ - for near-planar, near-circular 
tadpole orbits which evaluates to $0.065 \epsilon$ for Ceres and $0.053 \epsilon$ for Vesta.

The significant size of the sample of coorbitals under study prompted a search for trends in the distribution of 
$\lambda_{r}$. Fig.~\ref{fig:histo_lambda} shows a histogram of this quantity at $t=0$ for all current 
coorbitals and with a bin size of $30^{\circ}$. On this we superimpose, as a dashed line, a histogram of all objects
which are not currently coorbiting with either Ceres or Vesta. The vertical line segments indicate 
square-root Poissonian uncertainties. There appear to be no features that stand out above the uncertainties. 
Hence the coorbital resonance does not measurably affect the phasing of the populations of current coorbitals 
with respect to their secondary in this case.  
%
%
%
%
%
%

\section{Analysis of the Dynamics}
The dynamical context presents some similarities with coorbitals of the Earth and Venus such as non-negligible 
eccentricities and inclinations. Here we attempt to model the evolution using the framework of the restricted 
three body problem where a particle's state evolves under the gravity of the Sun and the secondary mass 
\citep{Namouni.et.al1999}. This is done through the expression
\begin{equation}
a^{2}_{ r}= C - \frac{8 \mu}{3} S \left( \lambda_{ r}, e, I, \omega \right)
\label{zvc}
\end{equation}
where 
\begin{equation}
S=\frac{1}{2 \pi}\int_{-\pi}^{\pi}\left( \frac{a_{S}}{|\mathbf{r} - \mathbf{r_{S}}|} - \frac{\mathbf{r} \cdot \mathbf{r_{S}}}{a \mbox{ }a_{S}} \right)  d \lambda .\label{exps}
\end{equation} 
Here, $a$, $e$, $I$, $\omega$ and $\lambda$ denote the semimajor axis, eccentricity, inclination, argument 
of pericentre and mean longitude of the particle's orbit. The subscript ``S'' is used to denote the same 
quantities for the secondary. The heliocentric position vectors of the particle and the secondary are 
denoted as $\mathbf{r}$ and  $\mathbf{r_{\rm S}}$ respectively. The relative elements $a_{r}$ and 
$\lambda_{r}$ are defined as
\begin{equation}
a_{r}= \left(a - a_{S}\right)/a_{S}\mbox{, } \lambda_{r}=\lambda - \lambda_{S}
\label{arlr}
\end{equation}
and $\mu$ is the mass of the secondary scaled to the total system mass. Eq.~\ref{zvc} can be seen as a 
conservation law where $C$ is the constant ``energy'' of the particle,  $a^{2}_{ r}$ its kinetic energy 
and the term containing $S$ its potential energy. \citeauthor{Namouni.et.al1999} showed that, as the left-hand 
side of this expression cannot be negative, it restricts, in general, the evolution of 
$\left(a_{r}, \lambda_{r}\right)$. A collision $\left(\mathbf{r} = \mathbf{r_{S}}\right)$ can only occur 
for specific combinations of values for $e$, $I$ and $\omega$. Hence, actual collisions are rare and the above 
formulation is generally valid.

\begin{figure}
\centering
\includegraphics[width=10cm,angle=90]{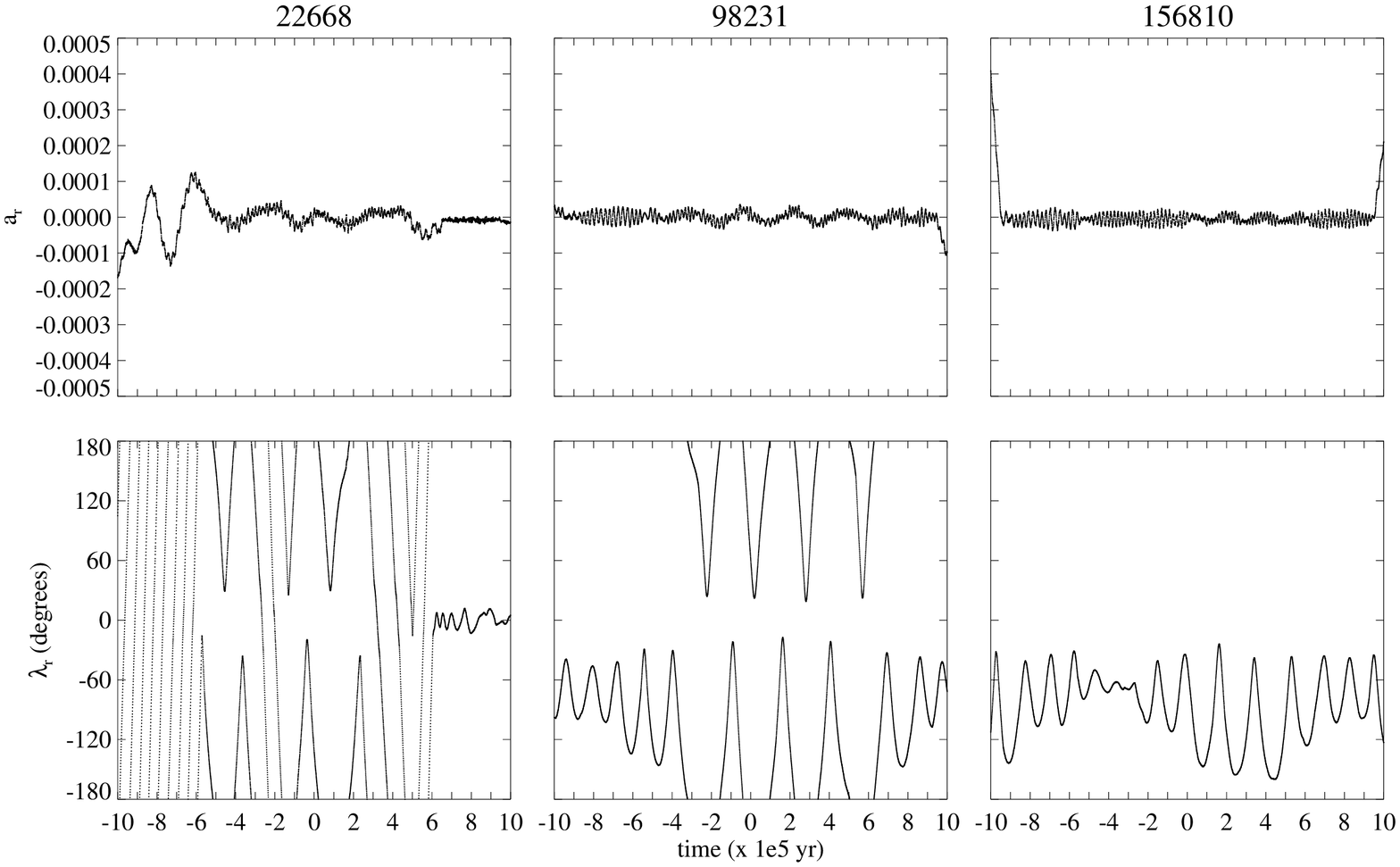}
\caption[As Fig.~\ref{fig:ceres_expls} but for coorbitals of Vesta.]{Christou and Wiegert 2010, Coorbitals of Ceres and Vesta}
\label{fig:vesta_expls}
\end{figure}

\begin{figure}
\vspace{-1cm}
%
%
\includegraphics[width=14cm,angle=0]{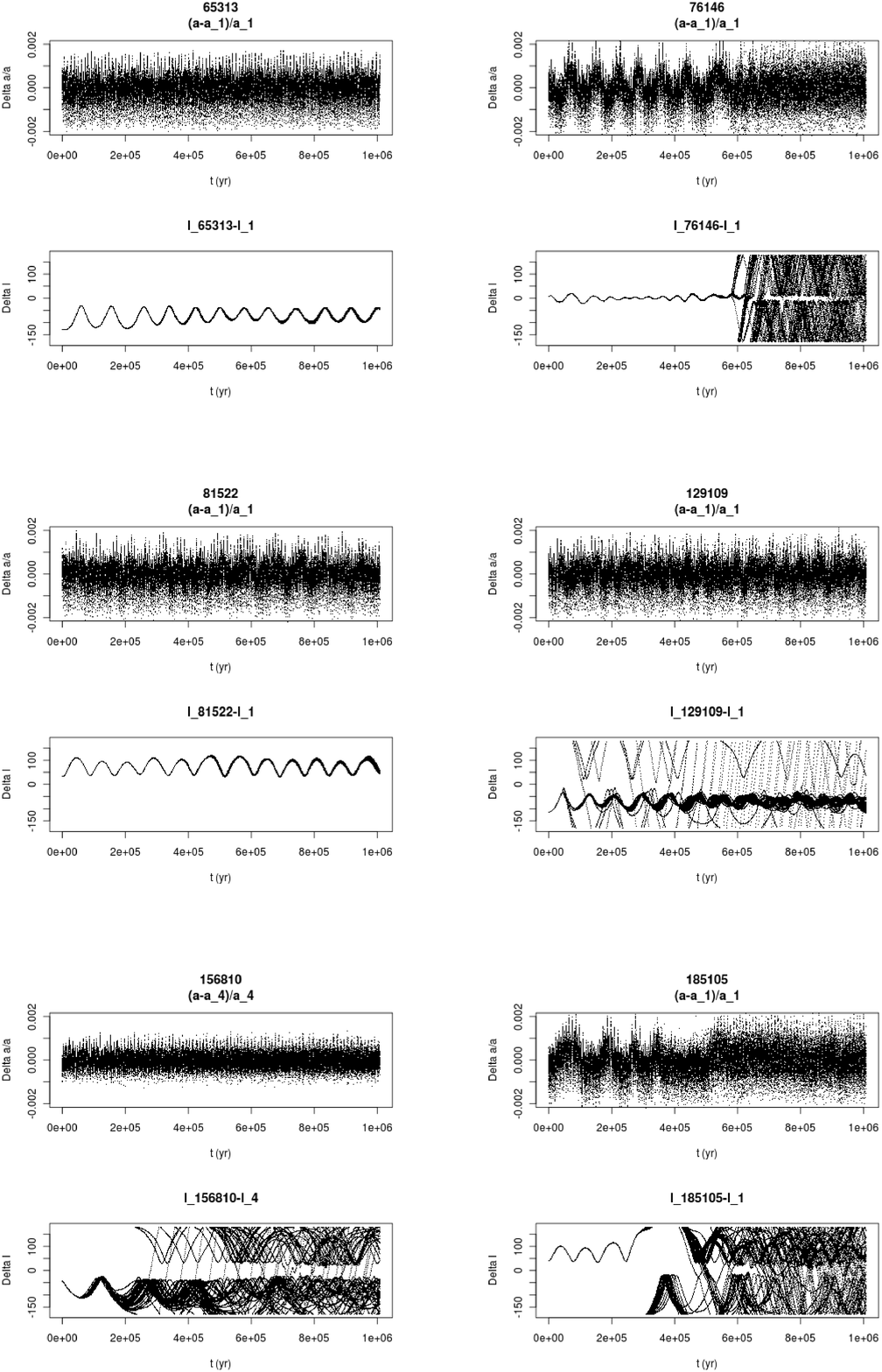}
\caption[Dynamical evolution of 600 clones of asteroids 65313 (Ceres coorbital; top left), 
76146 (Ceres coorbital; top right), 81522 (Ceres coorbital; centre left), 129109 
(Ceres coorbital; centre right), 156810 (Vesta coorbital; bottom left) and 185105 
(Ceres coorbital; bottom right) for $10^{6}$ yr as explained in the text.]{Christou and Wiegert 2010, Coorbitals of Ceres and Vesta}
\label{fig:stability}
\end{figure}

\begin{figure}
\centering
\includegraphics[width=10cm,angle=0]{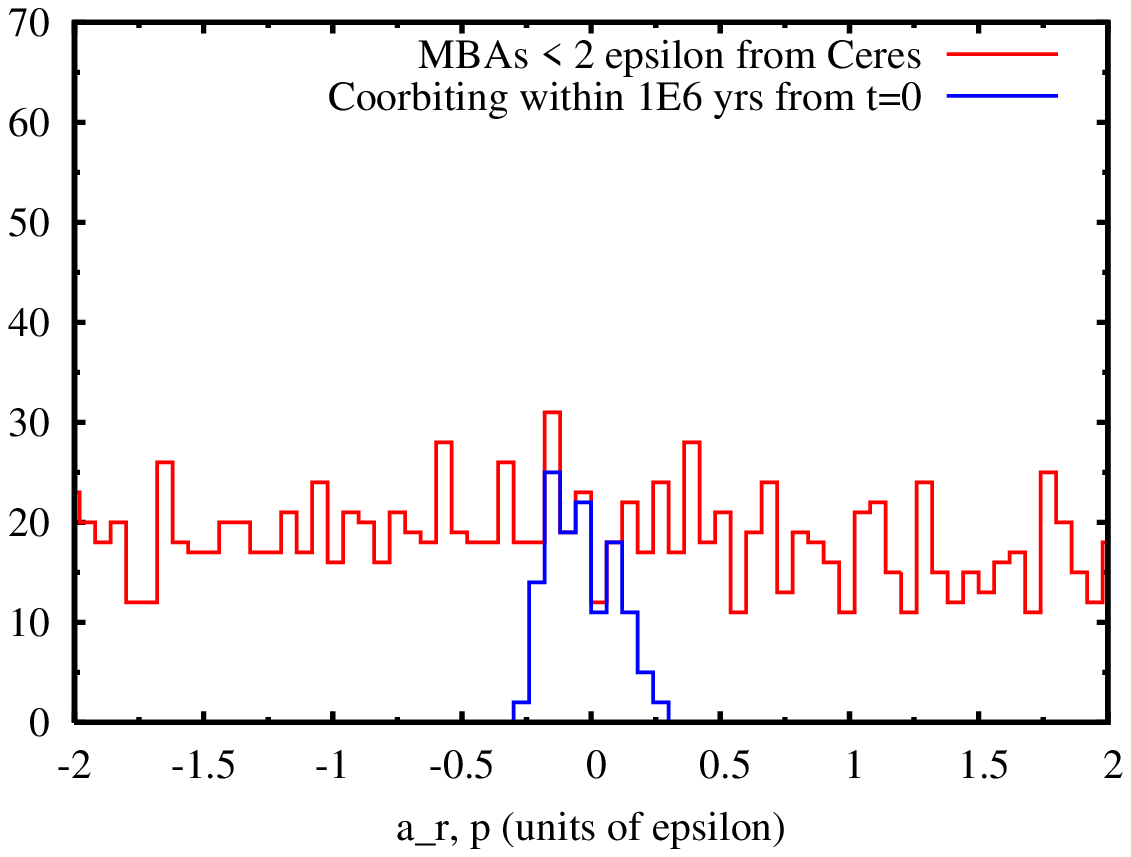}
\includegraphics[width=10cm,angle=0]{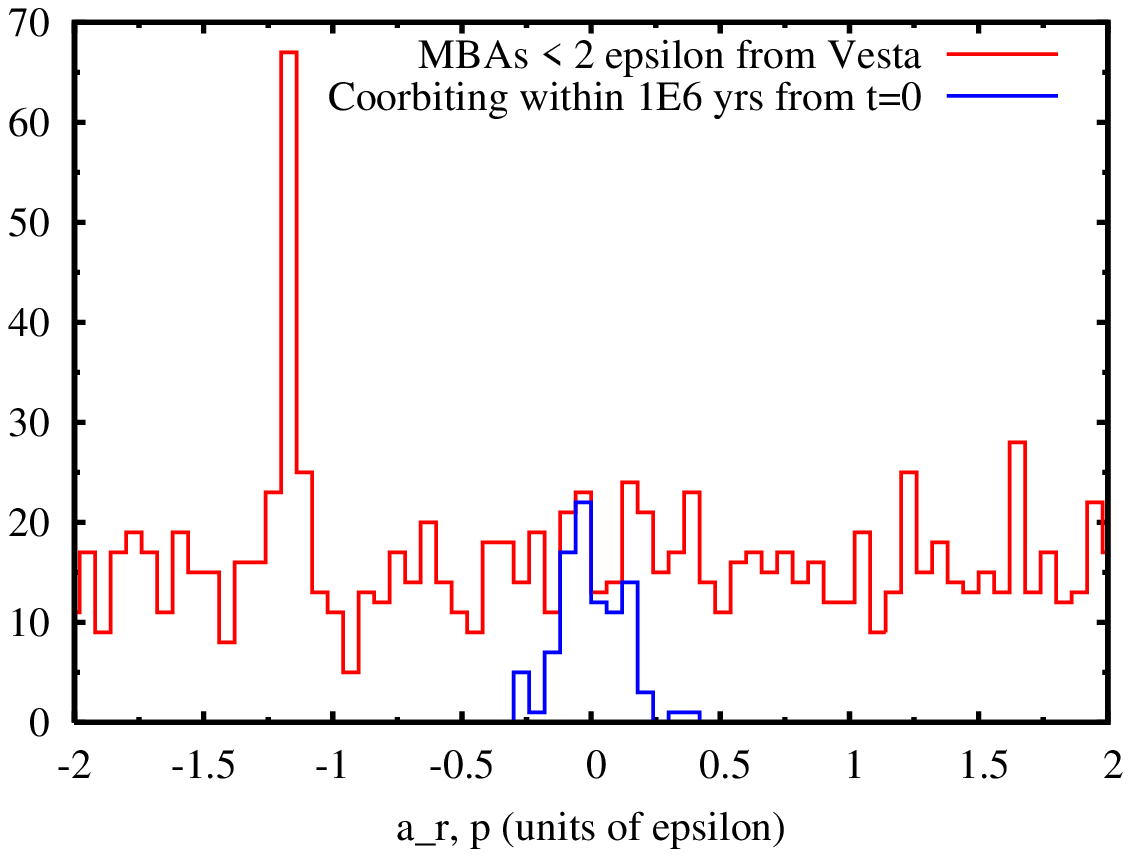}
\caption[Histograms of the statistical distribution of co-orbital asteroids of Ceres (upper panel) and Vesta 
(lower panel) within the coorbital regions of these massive asteroids as quantified by the relative proper
semimajor axis $a_{r\mbox{, }p}$.]{Christou and Wiegert 2010, Coorbitals of Ceres and Vesta}
\label{fig:histo_ap}
\end{figure}

\begin{figure}
\centering
\includegraphics[width=10cm,angle=0]{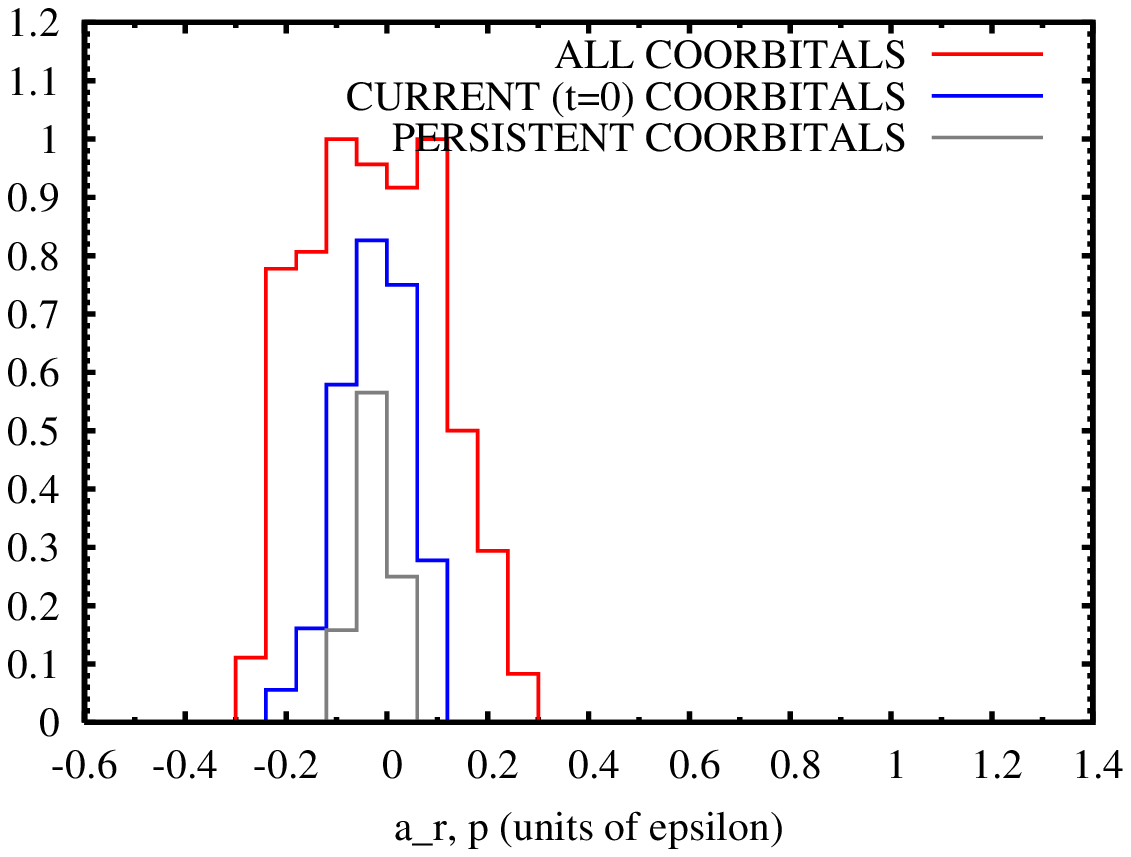}
\includegraphics[width=10cm,angle=0]{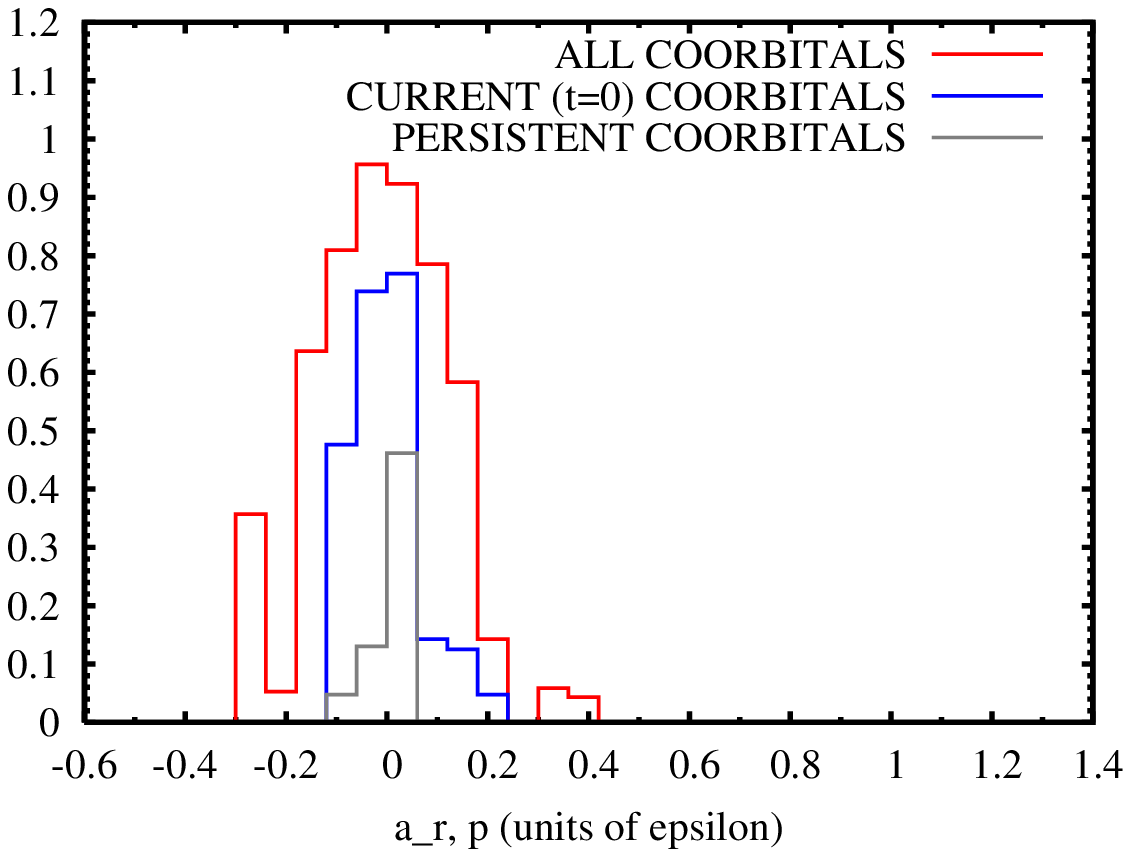}
\caption[Histograms of the statistical distribution of different types of Ceres (top) and Vesta (bottom) co-orbital asteroids divided over the total number of asteroids in each bin. The x-axis is in the same units as Fig.~\ref{fig:histo_ap}.]{Christou and Wiegert 2010, Coorbitals of Ceres and Vesta}
\label{fig:histo_ap2}
\end{figure}

\begin{figure}
\centering
\includegraphics[width=7cm,angle=0]{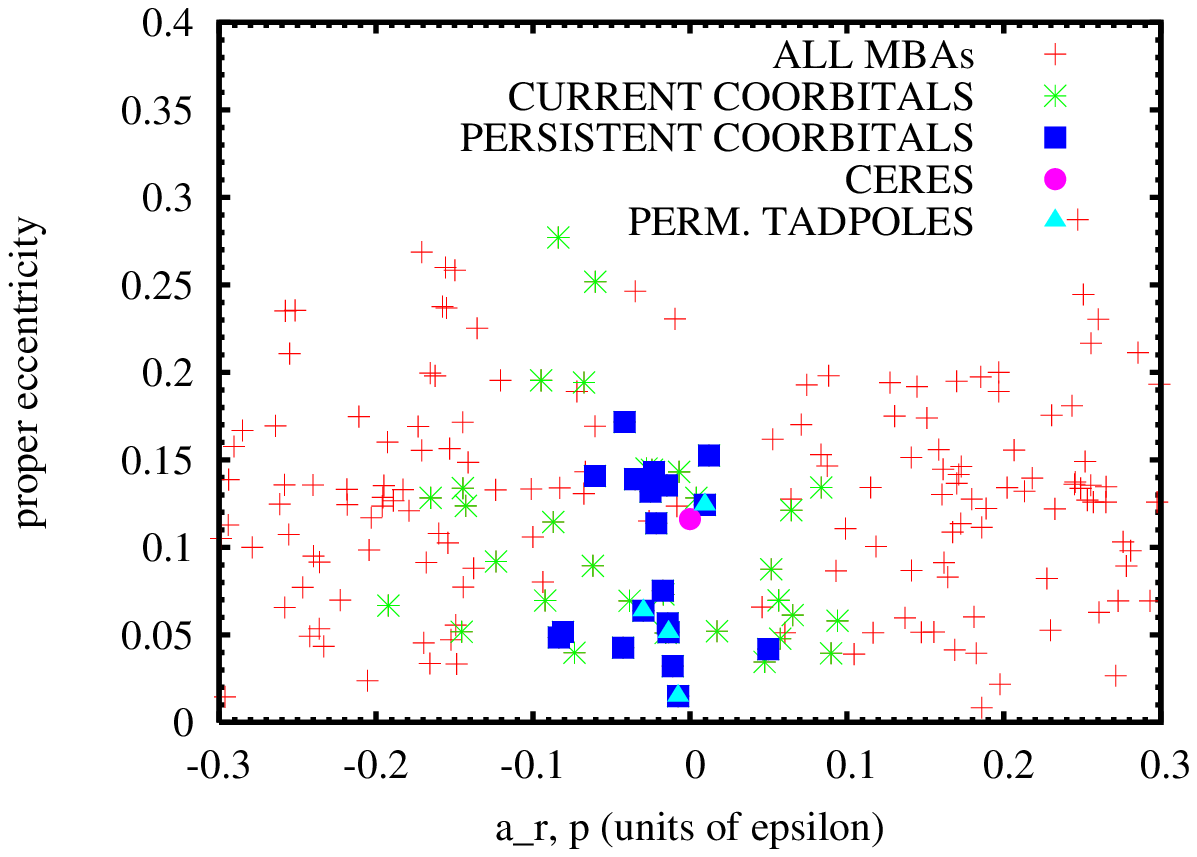}\includegraphics[width=7cm,angle=0]{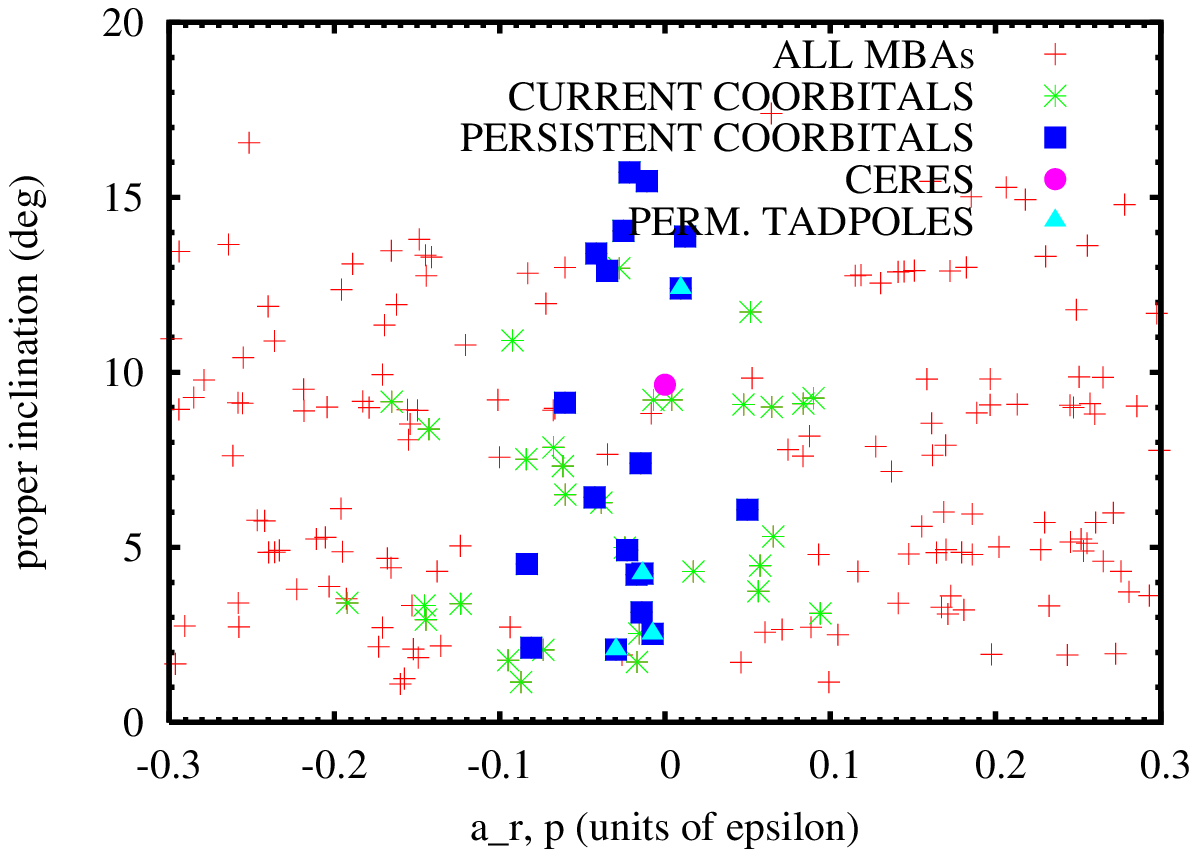}
\includegraphics[width=7cm,angle=0]{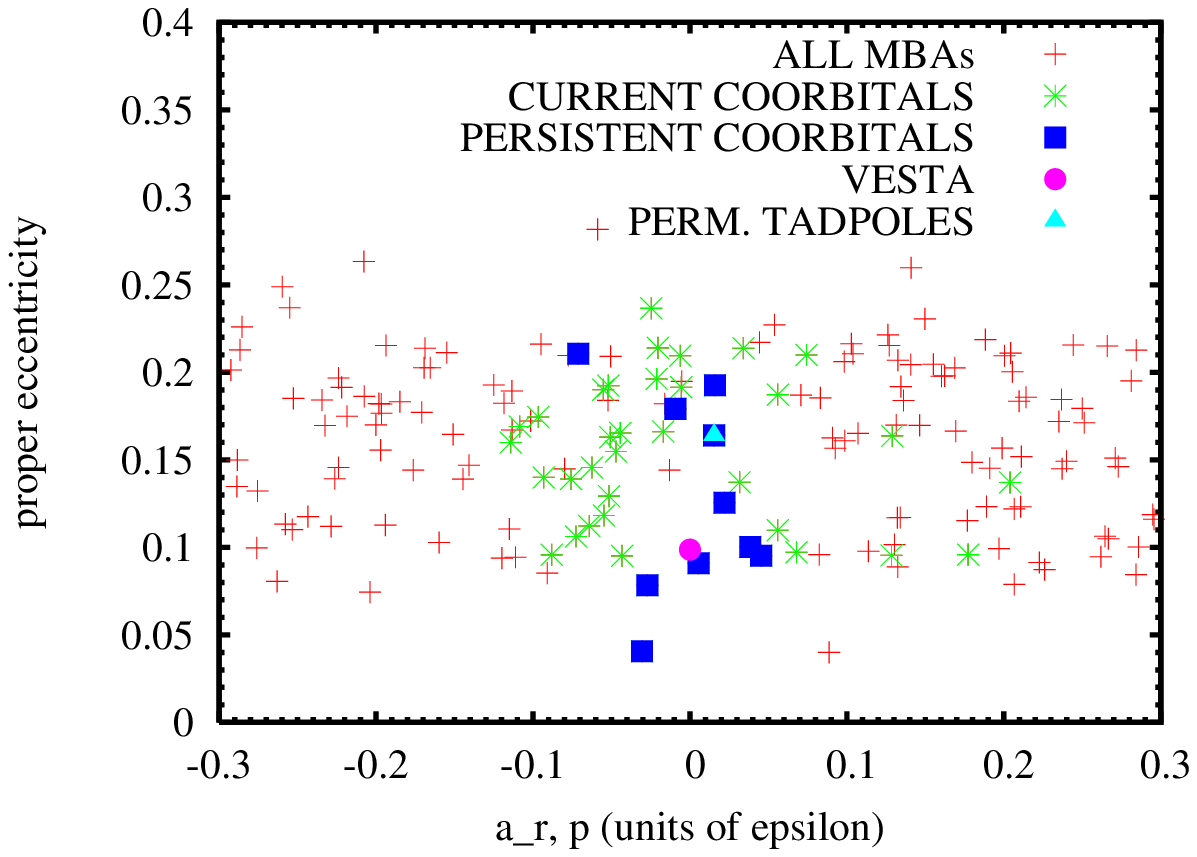}\includegraphics[width=7cm,angle=0]{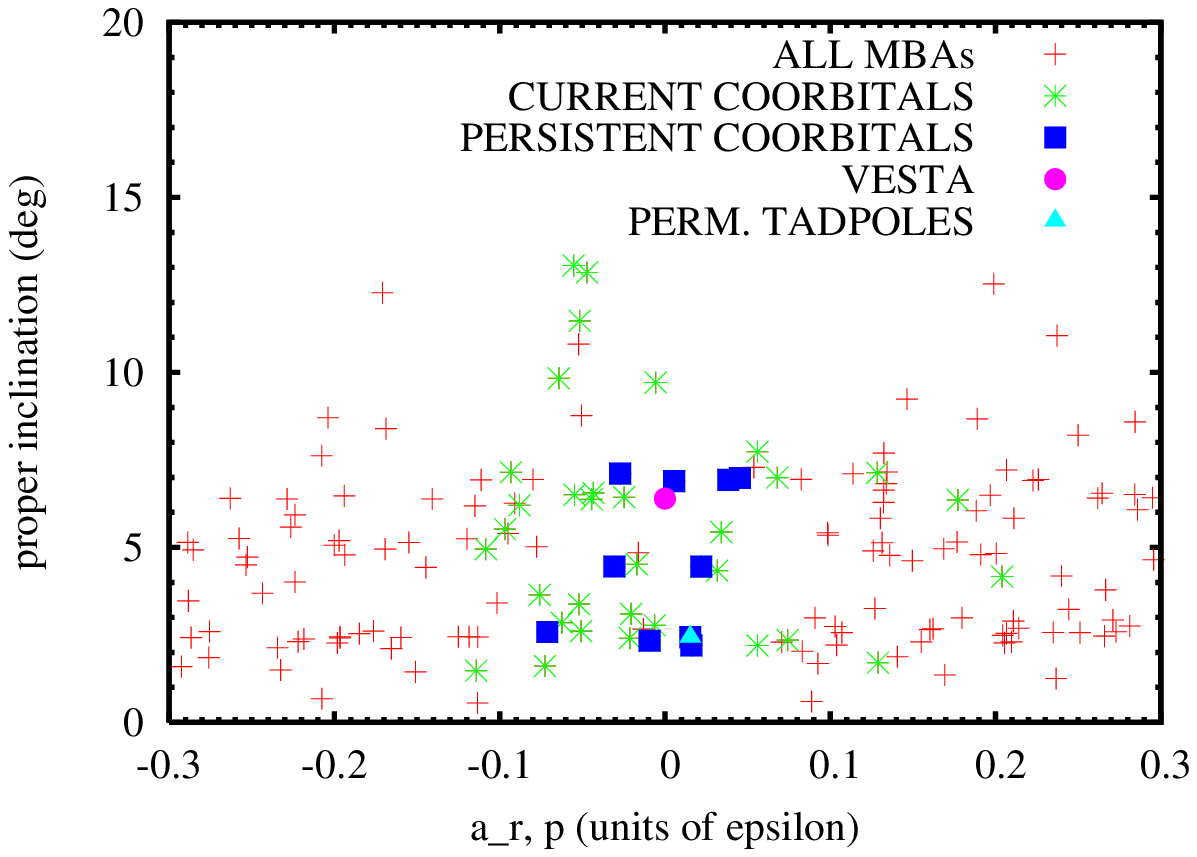}
\caption[Proper element distribution of different types of Ceres (top) and Vesta (bottom) co-orbital asteroids. The two left panels show the proper eccentricity as a function of the relative proper 
semimajor axis (units of $\epsilon$ as in Fig.~\ref{fig:histo_ap}) while the right two panels show the 
proper inclination.]{Christou and Wiegert 2010, Coorbitals of Ceres and Vesta}
\label{fig:aei}
\end{figure}

\begin{figure}
\centering
\includegraphics[width=12cm,angle=0]{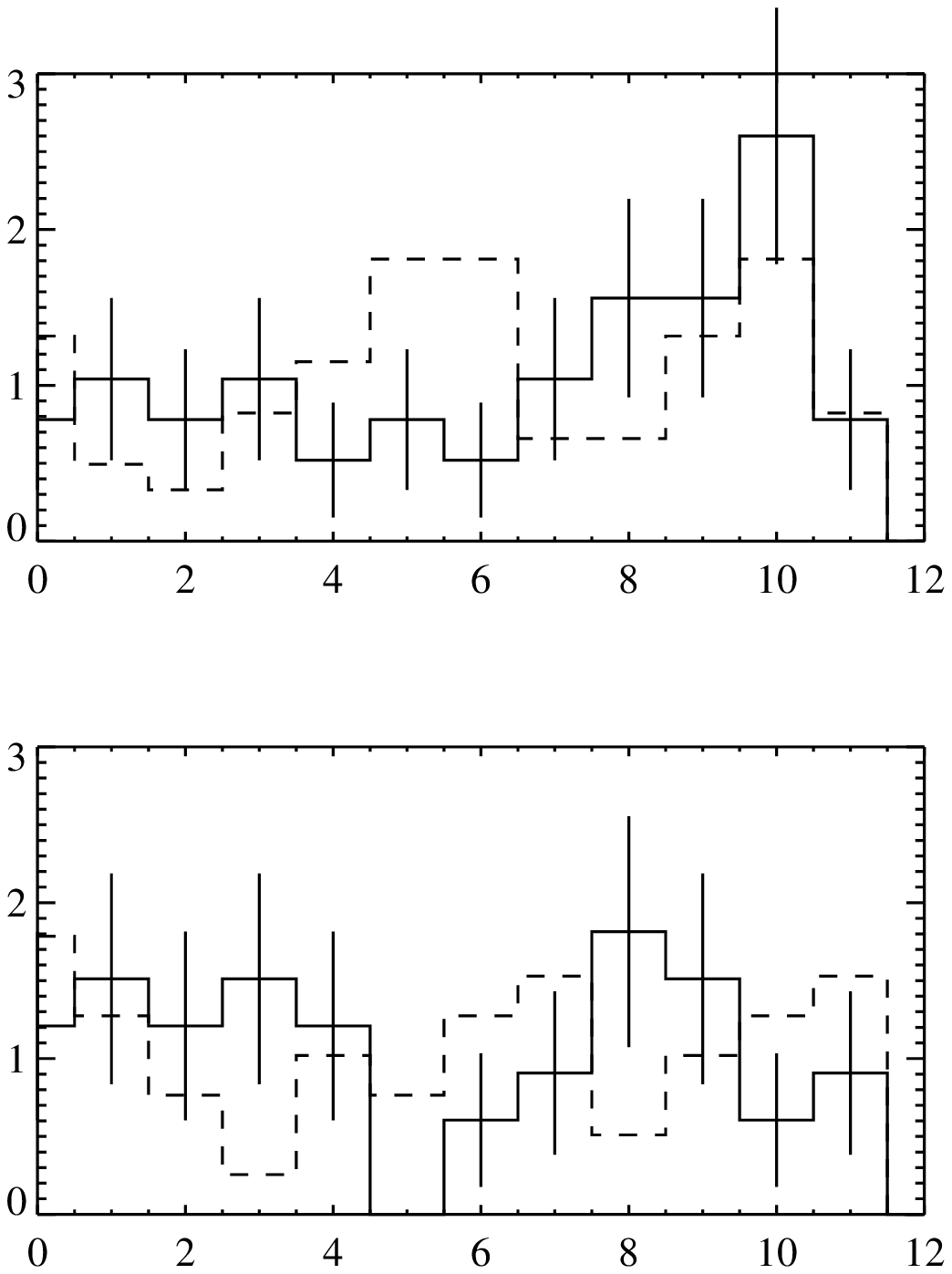}
\caption[Histogram of the relative longitudes $\lambda_{r}$ of asteroids currently coorbiting with Ceres 
(upper panel) or Vesta (lower panel). The position of the secondary 
($\lambda_{r}=0$) is at bin 6. Counts have been normalised to the average value per bin.]{Christou and Wiegert 2010, Coorbitals of Ceres and Vesta}
\label{fig:histo_lambda}
\end{figure}

For computational purposes, Eq.~\ref{exps} may be evaluated using standard two-body formulae 
\cite[eg][]{MurrayDermott1999} as $\dot{\lambda} \gg$ $\dot{e}$, $\dot{I}$, $\dot{\omega}$ and $\dot{\lambda}_{r}$. 
One other consideration that is specific to this paper is that the high frequency harmonics of $a$ are 
external to the R3BP (hence external to the model) and must somehow be removed before the above expressions 
may be used. However, at $t=0$, the relative proper semimajor axis $a_{r\mbox{, }p}$ can be considered 
to be this low-pass filtered value of $a_{r}$ in the Sun - Ceres/Vesta - MBA problem 
(see also Section 2). Hence, the constant $C$ can be evaluated and the model can be readily applied. 
In Fig.~\ref{fig:model_fit} we show $S$ profiles for each of the MBAs in Figs~\ref{fig:ceres_expls} and 
\ref{fig:vesta_expls} compared to the location of the point $\left(\lambda_{r}\mbox{, }{\cal E}=3C/(8\mu)\right)$ 
at $t=0$ (dotted circle). As motion is restricted to the domain above $S$, the model predicts that the Ceres coorbitals 
71210, 81522, 76146 are currently in $\mbox{L}_{5}$ tadpole, $\mbox{L}_{4}$ tadpole and Quasi-Satellite 
libration respectively. Similarly the Vesta coorbitals 22668, 98231 and 156810 are predicted to be in horseshoe, horseshoe 
and $\mbox{L}_{5}$ libration respectively. Referring to the Figure, the model apparently succeeds in 5 out of the 6 cases, 
but fails in the case of the Ceres coorbital 71210 where the observed mode of libration is a horseshoe.

This is probably due to the fact that Eq~\ref{zvc} is evaluated when $a_{r}=0$ i.e. at the turning points of the libration.
In \cite{Namouni.et.al1999}, $\dot{\lambda}_{r} \gg \dot{e}\mbox{, }\dot{I}\mbox{, }\dot{\omega}$ and 
the orbit can be considered ``frozen'' during a libration cycle. In our case, however, we observe that 
$\dot{\omega} > \dot{\lambda_{r}}$ because of the small mass of the secondary. Incidentally, this parity in timescales
may also account for the general lack of compound libration modes for these coorbitals, as $\omega$ controls 
the relative height of the maxima of $S$ on either side of $\lambda_{r}=0$. 

To quantify the effect that this has to our model, we evaluated $\cal{E}$ against $S$ for the example 
MBAs shown in Figs~\ref{fig:ceres_expls} and \ref{fig:vesta_expls} but at different values of $\lambda_{r}$ and 
$\omega$. We found that determination of $S$ is generally insensitive to $\omega$, except near the local maxima 
bracketing the origin on the $\lambda_{r}$ axis. Physically, these correspond to the closest possible cartesian 
distances between the particle and the secondary so it is not a surprise that they are sensitive to the orbital 
elements. Particularly for the case where the model failed, ${\cal E} - S \simeq 0.6 $ when the asteroid reaches 
the far end of the model tadpole ($\lambda_{r} \sim - 130^{o}$) and the potential maximum at 
$\lambda_{r}=180^{o}$ i.e. the object is classified as a horseshoe in agreement with the numerical integrations. Hence, 
this method for determining the resonant mode is formally valid where the object is currently near the turning 
point of the libration i.e. those of MBAs 81522, 76146 and 156810. In the other cases, the more involving process 
of monitoring the quantity  ${\cal E} - S $ in the integrations for a time period comparable to a libration cycle 
would be necessary to establish the libration mode. 

Finally, we wish to understand the stability of the QS librator of Ceres, 76146, in the context of 
our findings. The sensitivity of $S$ to $\omega$ for the local maxima near $\lambda_{r}=0$ is important 
as these features of the potential are the ``gatekeepers'' for evolution in and out of the QS. This object 
was captured as a QS from a passing orbit and completes 7 cycles in this mode before becoming a horseshoe. 
It is currently at $\lambda_{r}\simeq 8^{o}$. Irrespective of the value of ${\cal E}$, escape from the QS 
is certain when $S_{t=0} \leq S_{\lambda_{r}=0}$ i.e. when the extremum at the origin becomes a local maximum. 
According to the model this occurs for $-34^{0} < \omega < +12^{o}$. Keeping in mind the $\pi$ periodicity 
of $S$ in $\omega$, the asteroid will turn back $ 360/2/46 \sim 4 $ times before it escapes, in good agreement with what is observed in the 
numerical simulation. 

\section{The role of other massive asteroids}
\citet{Christou2000a} showed that Venus and Mars play a key role in the evolution of Earth co-orbitals. 
These can force transitions between different libration modes or escape from the resonance altogether. 
Here the only candidates available to play a similar role are other massive asteroids. In this paper we have 
focused on the effects of Ceres and Vesta - as well as Pallas - on Vesta or Ceres co-orbitals respectively.
In a first experiment to determine their role (if any), we have integrated the same six asteroids as in 
Section 3 but, in the first instance leaving Pallas out of the model (``No Pallas'' or NP) and, in the 
second, only under the gravity of the secondary (ie Ceres or Vesta as appropriate; ``Secondary Only'' or SO). 
We find that the evolution of the Ceres Trojans 65313 and 81522 and of the Quasi-Satellite 76146 are the same 
in both of these runs as well as the original runs where all three massive asteroids were present (``All Masses'' 
or AM). In contrast, we find significant differences in the evolution of the remaining three asteroids, 129109, 
156810 and 185105 (Fig.~\ref{fig:encounters}). In the SO runs (top row), all clones of these Trojan librators persist as 
such. This is not the case for the NP and AM runs (middle and bottom rows respectively). 
Most clones ultimately leave this libration mode, with the exception of 129109 in the AM runs. 
Although it is difficult, on the basis of a sample of six, 
to attempt to decouple the effects of the individual massive asteroids, these results indicate 
that their presence does have an effect on the lifetime of the co-orbital configurations we have found.   
   
To quantify these in a more statistically robust sense, we used MERCURY as before to re-integrate the 
original 129 and 94 co-orbital MBAs of Ceres and Vesta respectively under the SO, NP and AM models for 
$10^{6}$ yr in the future (ie half the timespan of the original integrations). We then check which MBAs survive, 
either as ``persistent'' or ``single mode'' libration in the new runs. The results are shown in Table~\ref{tab:stability}. 
As far as persistency is concerned, the fraction of those co-orbitals in the respective samples is between 
20\% and 25\%  for the three models examined (SO, NP and AM) but also for both Ceres and Vesta co-orbitals. 
Upon closer inspection, we find that approximately the same number of persistent co-orbitals (16 for Ceres and 
14 for Vesta) are common between the three models. Thus, Vesta harbours a slightly larger fraction of these 
co-orbitals than does Ceres. The remainder of the persistent population is composed of objects that gain or 
lose the persistency property from one model to the next. There is no apparent trend towards one direction 
or the other; objects that lose this property are replenished by the objects which gain it. Thus, it seems that 
persistency of co-orbital motion is independent of the adding or removing massive asteroids from the model. 
If this premise is correct, then a mechanism that can explain the observations is the intrincic chaoticity 
of the orbits. We saw in Section 3 that the amount of chaotic `noise' in the proper semimajor axis is 
comparable to the width of the coorbital zones of Ceres and Vesta. It is thus reasonable to expect that 
some objects are removed from the resonance while others are injected into it as a consequence of that 
element's random walk. 

The behaviour of single-mode coorbital libration is altogether different. For Ceres, we find that the 
addition of other massive asteroids increases the number of persistent single-mode librators, from 6 (SO) 
to 14 (NP, AM). For Vesta, we observe the opposite trend: adding Pallas and Ceres to the model decreases 
the number of such librators from 11 (SO) to 3 (NP) and 5 (AM). In addition, only one object, 87955, 
persisted as a Ceres $\mbox{L}_{4}$ Trojan in all three models. The statistics are marginally significant for these 
low counts. Nevertheless, they compelled us to explore possible causes.

It is tempting at this point to try and correlate persistency with the objects' Lyapounov Characteristic 
Exponent (LCE), an established quantifier of chaoticity, available from the proper element database used in this paper. 
However, we point out that these LCEs were computed under a dynamical model that does not involve any 
massive asteroids. As it is not clear how the presence of the co-orbital resonance will affect the 
determination of the LCE, we refrain from attempting such a correlation in this work.  

 In Fig.~\ref{fig:nencceres} and \ref{fig:nencvesta} we present the distribution of the number of Ceres or Vesta 
coorbitals respectively that undergo a given number of encounters within 5 Hill radii ($R_{H}=a \epsilon$) of the massive 
asteroids in the three models. A number of interesting features are evident. Firstly, persistent coorbitals 
(green and blue columns), as well as a certain fraction of non-persistent coorbitals (red columns; 25-30\% 
for Ceres coorbitals, $\sim$50\% of Vesta coorbitals), do not, in general, approach the secondary. This feature 
was also identified in \citet{Christou2000a} for Earth co-orbitals (cf Fig.~8 of that paper). It seems to be 
a generic feature of coorbital dynamics at high eccentricity and inclination orbits although the cause-and-effect relationship is not yet clear.

In addition, the distribution 
of encounters for non-persistent coorbitals (red) shows a tail; in other words, many of these objects undergo 
many encounters with the secondary. This is probably due to the slowness of the evolution of the relative longitude
of the guiding centre compared to the epicyclic motion. Indeed, we find that these encounters are not randomly 
distributed in time but occur in groups typically spanning a few centuries. In contrast, those related to 
massive asteroids other than the secondary have an upper cutoff (5 encounters for Pallas and Vesta,  10 for Ceres). 
The shapes of the distributions for the three classes of coorbitals are similar. The apparently high frequencies 
observed in the case of Ceres encounters for single-mode persistent coorbitals of Vesta are probably due to the 
small size of the population in those classes. The better populated distributions in the AM and SO models 
mimic the distributions of the other two classes we investigated.

A different way to look at the data is to create histograms of minimum distances, since distance is one of 
the factors (velocity being the other) that determine the magnitude of the change in an MBA's orbit. 
Fig.~\ref{fig:nencdistceres} and \ref{fig:nencdistvesta} show the distribution of these distances in units 
of $R_{\rm H}$. Bin $i$ contains all recorded encounter distances between $\left(i - 1 \right)R_{\rm H}$ 
and $i R_{\rm H}$ and have been normalised with respect to the area of the corresponding annulus of width 
$R_{\rm H}$ on the impact plane. For Ceres coorbitals we do not discern any statistically significant variations, in other words 
the counts within the different bins are the same given the uncertainties. For Vesta coorbitals, the situation is similar 
with one exception: we note that single mode persistent coorbitals (blue) do not approach Ceres closer
than one Hill radius. Due to the low counts, we cannot exclude the possibility that we are looking at 
statistical variation in the data. However, if it is the signature of a real trend, it would mean that 
encounters with Ceres may cause Vesta coorbitals to exit a particular libration mode.

Regardless of their significance at population level, there is clear evidence in the data that Ceres encounters 
can affect the orbital evolution of coorbitals of Vesta. Fig~\ref{fig:encceres} shows two instances where 
Vesta coorbital MBAs 45364 and 164791 leave the resonance following close encounters with Ceres deep within 
the Hill sphere of that dwarf planet. We have searched for similar occurences in the evolution of Ceres coorbitals but
without success. Neither Pallas nor Vesta appear capable of playing a similar role. This is probably due to their smaller 
masses (hence physically smaller Hill spheres) but also, specifically for the case of Pallas, the higher encounter velocities.

\section{Quasi-Satellites}
This Section is devoted to the existence, as well as stability, of so-called ``Quasi-Satellite'' (QS) or 
``bound'' orbits. These appeared first in the literature as Retrograde Satellite (RS) orbits 
\cite[as it turned out, a special case of the QS state;][]{Jackson1913} and later studied in the context 
of dynamical systems analysis \citep{Henon1969,HenonGuyot1970}. More recently, the survival of this libration 
mode in the real solar system was examined by \citet{MikkolaInnanen1997} and \citet{Wiegert.et.al2000}. 
Currently, two known Quasi-Satellites exist for the Earth \citep{Wajer2010} and one for Venus \citep{Mikkola.et.al2004}.

Several instances of QS libration relative to Ceres and Vesta were found in our simulations. Before these are 
discussed in detail, we introduce some elements of a convenient theoretical framework to study QS motion 
dynamics, namely that by \citet[][see also Hen\'{o}n and Petit, 1986]{Namouni1999}. It employs the set of 
relative variables
\begin{eqnarray}
x_{r}&=&e_{r} \cos \left(nt -\varpi_{r}\right) + a_{r}\\
y_{r}&=&- 2 e_{r} \sin \left( n t -\varpi_{r} \right) + \lambda_{r}\\
z_{r}&=&I_{r} \sin \left( n t - \Omega_{r} \right)
\end{eqnarray} 
where $n$ is the mean motion and $e_{r}$, $I_{r}$, $\varpi_{r}$
and $\Omega_{r}$ the relative eccentricity, relative inclination, relative longitude of pericentre
and relative longitude of the ascending node as defined in \citet{Namouni1999} respectively. In this formulation, 
the motion is composed of the slow evolution of the guiding centre $\left(a_{r}\mbox{, }\lambda_{r}\right)$ to 
which a fast, three-dimensional epicyclic motion of frequency $n$ and amplitude proportional to $e_{r}$ and $I_{r}$ is superposed. Examples 
of guiding centre libration while in QS mode have been shown in Figures~\ref{fig:ceres_expls} (bottom right panel) and 
\ref{fig:vesta_expls} (bottom middle panel). To illustrate the relationship between the two components of 
the motion, we show in Fig~\ref{fig:quasisats} two examples of QS motion recovered from our simulations. 
The left panel shows the cartesian motion of MBA 50251 in a cartesian heliocentric frame that rotates with the mean motion of
Ceres (`C') and for a period of $\sim 5 \times 10^{4}$ yr. The guiding centre libration is indicated by the short arc straddling the 
secondary (actually a closed loop of width $\sim 10^{-3}$ in $a_{r}$) while the cartesian motion along 
the epicycle path appears as a loop. The inset shows the evolution of the relative longitude $\lambda_{r}$ 
for $10^{6}$ yr including the period of libration around $0^{\circ}$. The example on the right panel shows 
the motion of MBA 121118 in a frame rotating with the mean motion of Vesta (`V') during a period of QS libration 
around Vesta lasting for $7 \times 10^{5}$ yr.
As in the previous case, the guiding centre libration is indicated by the short arc straddling the 
secondary. Here, the higher amplitude of the fast harmonics in the evolution 
of $a_{r}$ act to smear out the epicycle to some extent. It is also evident in these plots 
that QS librators are physically located well outside the secondary's Hill sphere and should not be 
confused with keplerian satellites.

\begin{figure}
\centering
\includegraphics[width=5cm,angle=0]{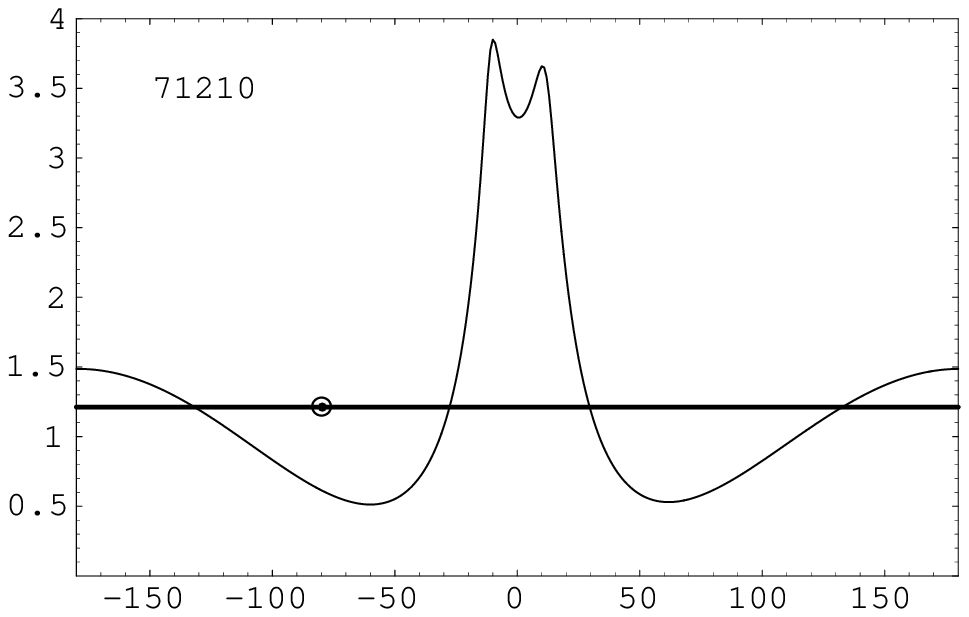}\includegraphics[width=5cm,angle=0]{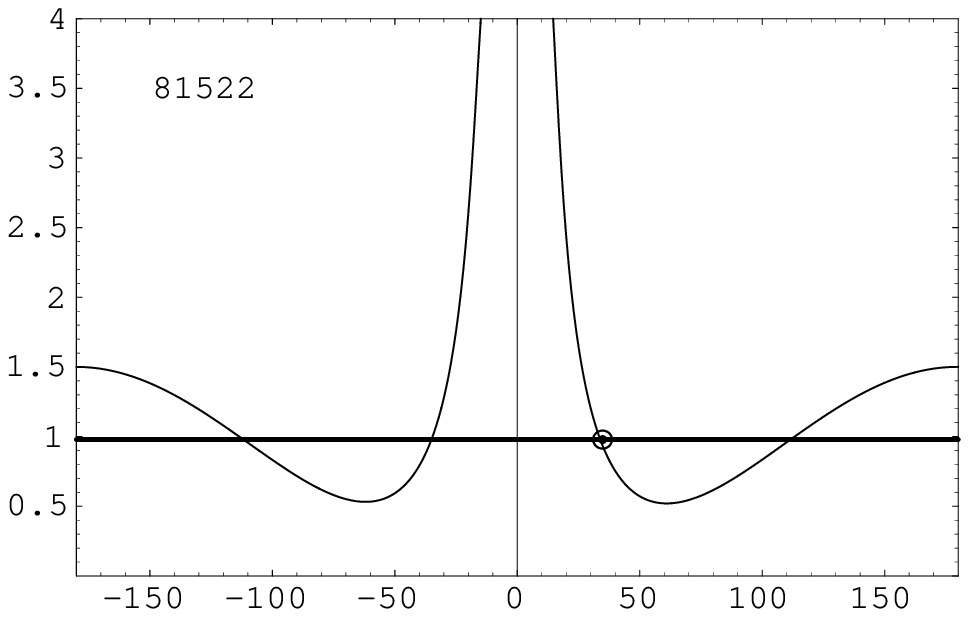}\includegraphics[width=5cm,angle=0]{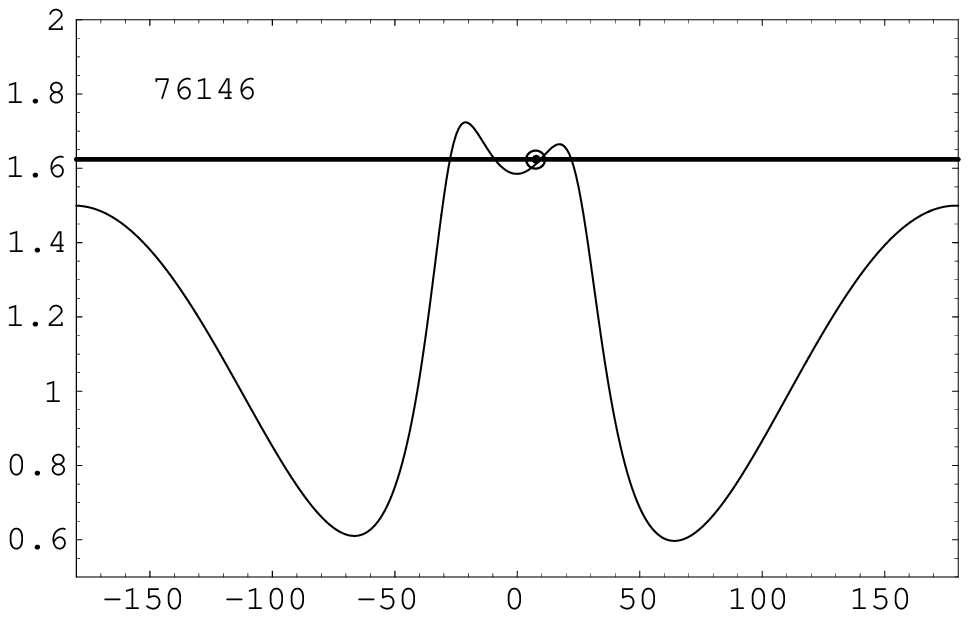}
\includegraphics[width=5cm,angle=0]{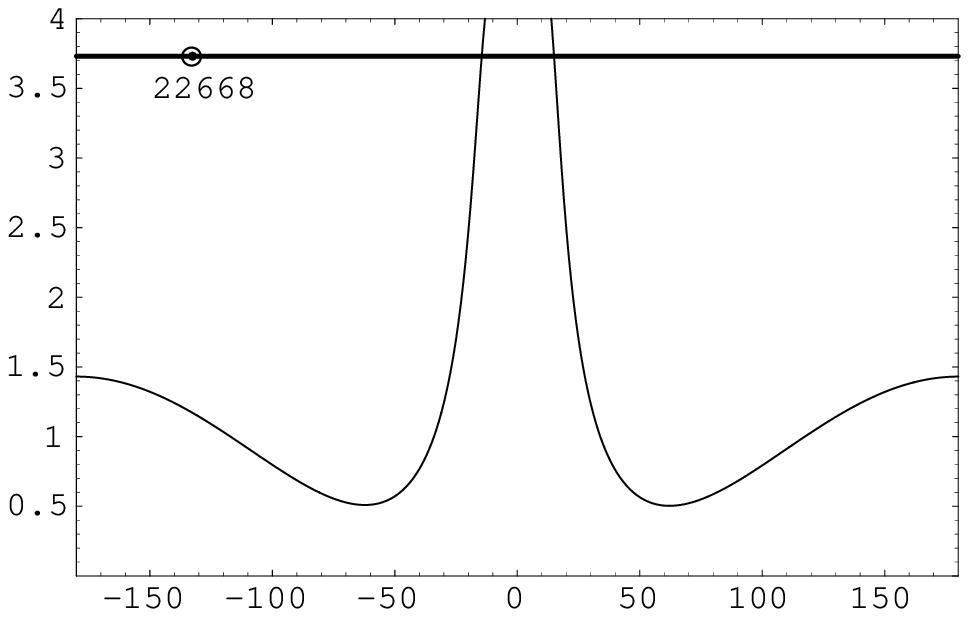}\includegraphics[width=5cm,angle=0]{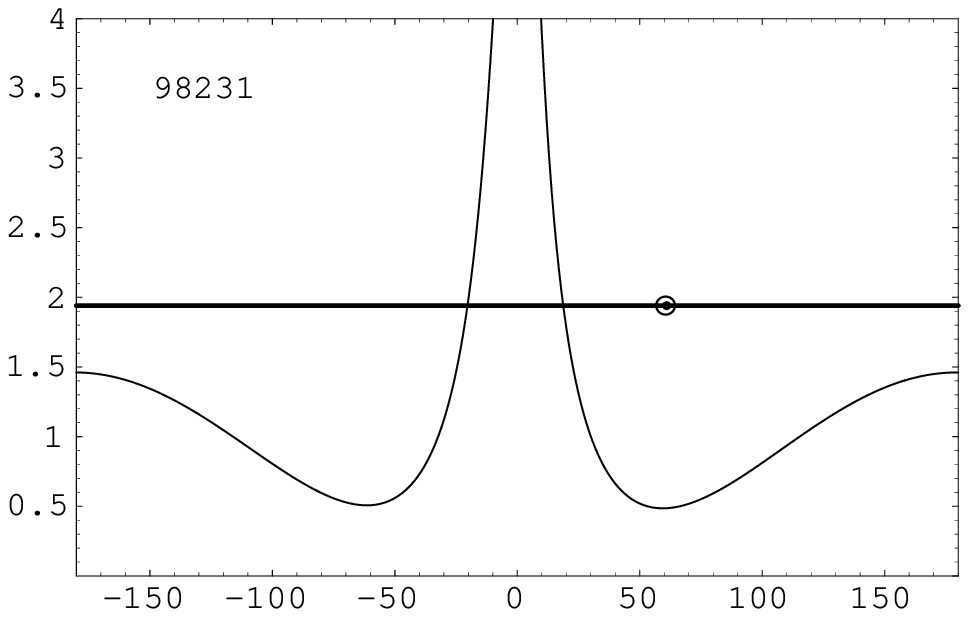}\includegraphics[width=5cm,angle=0]{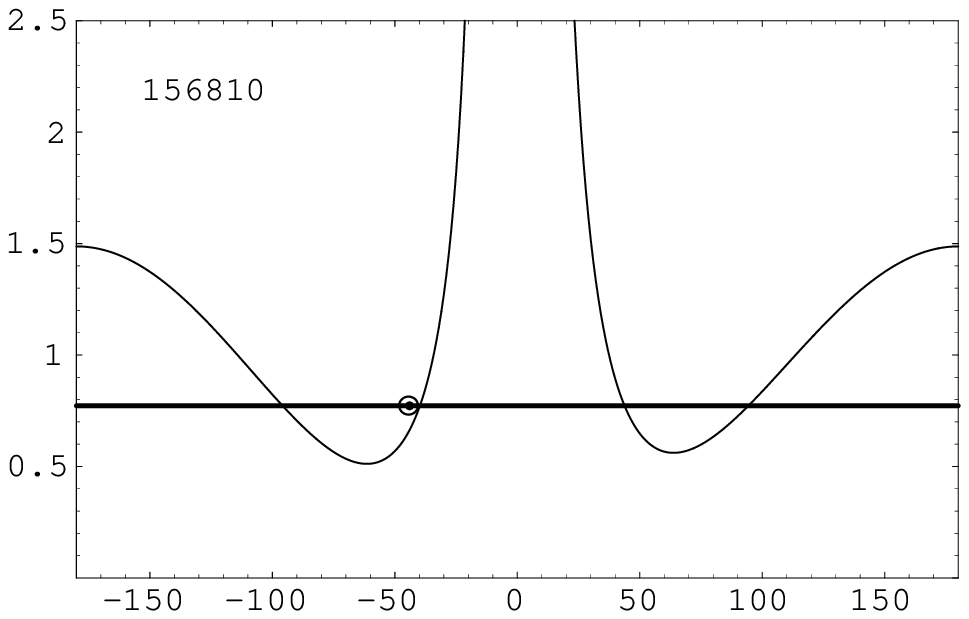}
\caption[Model fits to the motion of Ceres (top row) and Vesta (bottom row) coorbital MBAs shown in Figs~\ref{fig:ceres_expls} and \ref{fig:vesta_expls} respectively. 
The curve, horizonal line and dotted circle correspond to the profile of $S$ (Eq.~\ref{exps}), the quantity $3 C / 8 \mu$ and the point  $\left(\lambda_{r}\mbox{, }3 C / 8 \mu\right)$ respectively at $t=0$ as functions of the relative longitude $\lambda_{r}$ in degrees.]{Christou and Wiegert 2010, Coorbitals of Ceres and Vesta}
\label{fig:model_fit}
\end{figure}

\begin{figure}
\centering
\includegraphics[width=15cm,angle=0]{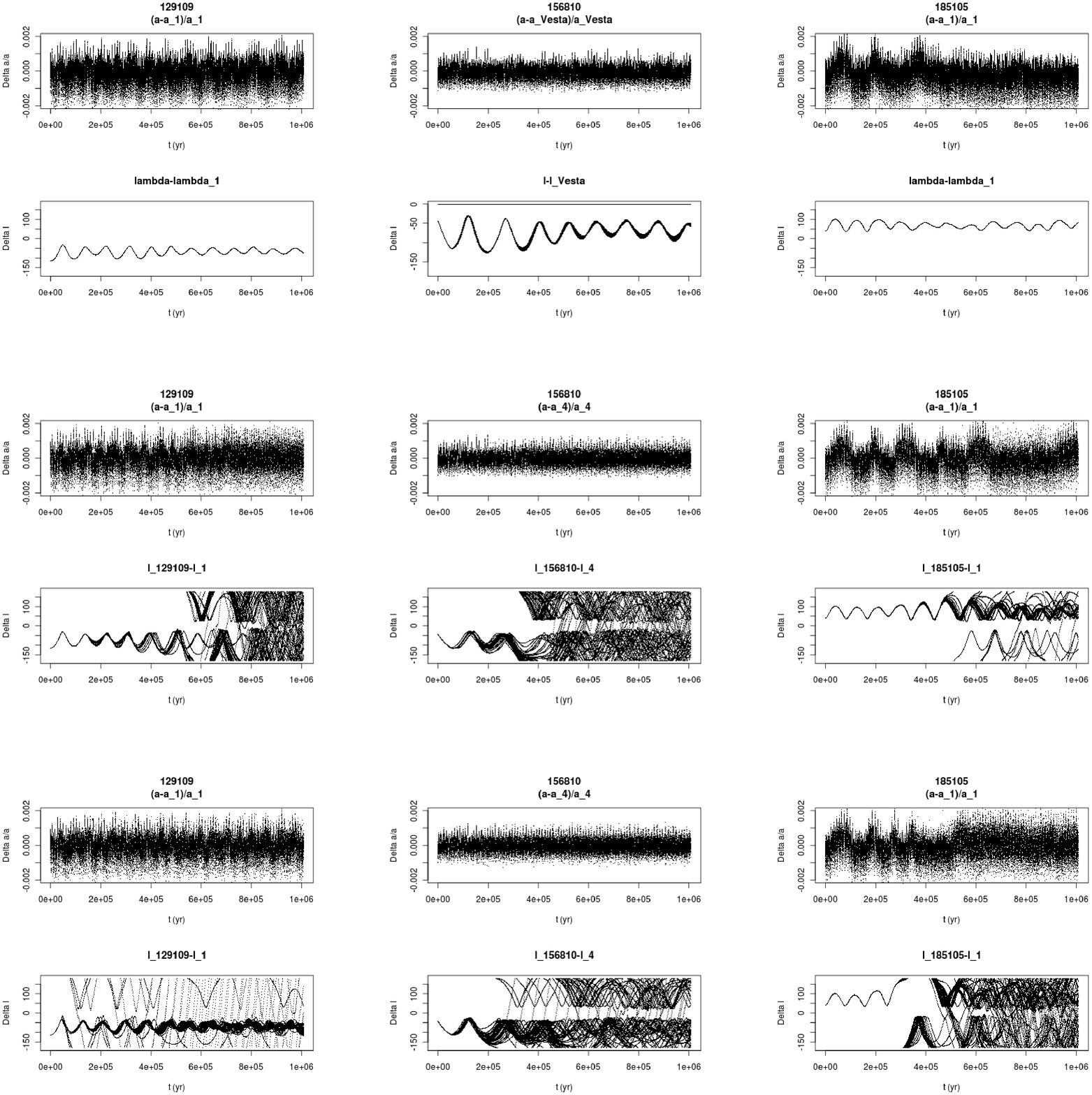}
\caption[Dynamical evolution of 300 clones of asteroids 129109, 156810 and 185105 (left, centre and middle column respectively) for the three models discussed in the text.]{Christou and Wiegert 2010, Coorbitals of Ceres and Vesta}
\label{fig:encounters}
\end{figure}

\begin{figure}
\centering
\includegraphics[width=15cm,angle=0]{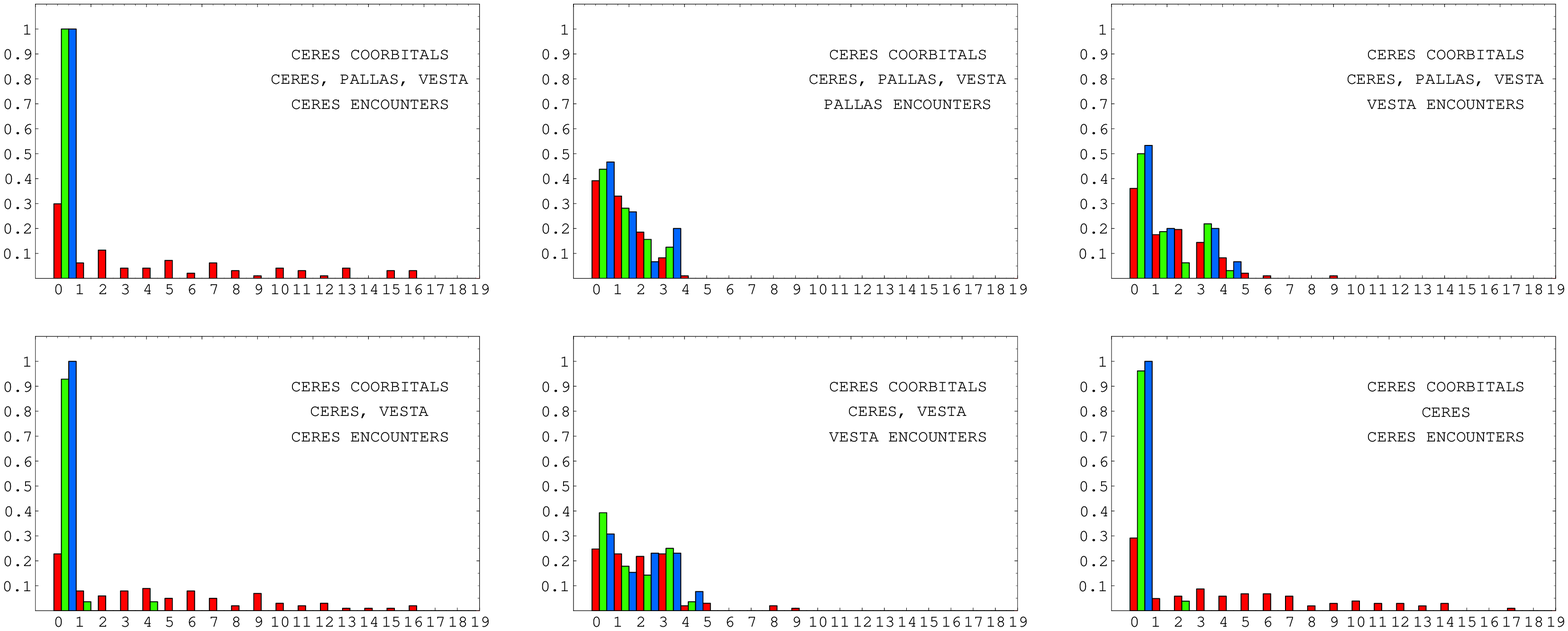}
\caption[Histograms of numbers of encounters ($< 5 R_{\rm H}$) of Ceres co-orbitals with massive 
asteroids in our three models. 
Bins have been normalised by the total number of objects in 
each persistency class.]{Christou and Wiegert 2010, Coorbitals of Ceres and Vesta}
\label{fig:nencceres}
\end{figure}

\begin{figure}
\centering
\includegraphics[width=15cm,angle=0]{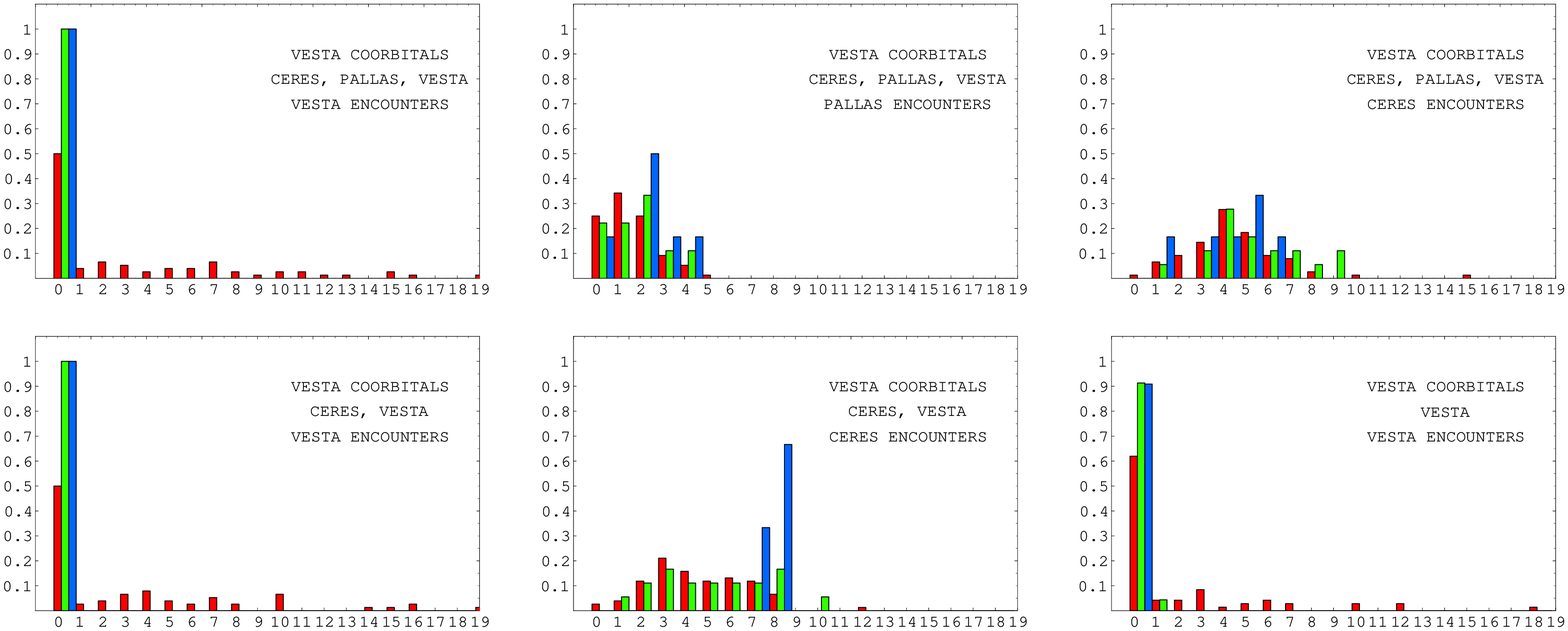}
\caption[As Fig.~\ref{fig:nencceres} but for Vesta co-orbitals.]{Christou and Wiegert 2010, Coorbitals of Ceres and Vesta}
\label{fig:nencvesta}
\end{figure}

\begin{figure}
\centering
\includegraphics[width=15cm,angle=0]{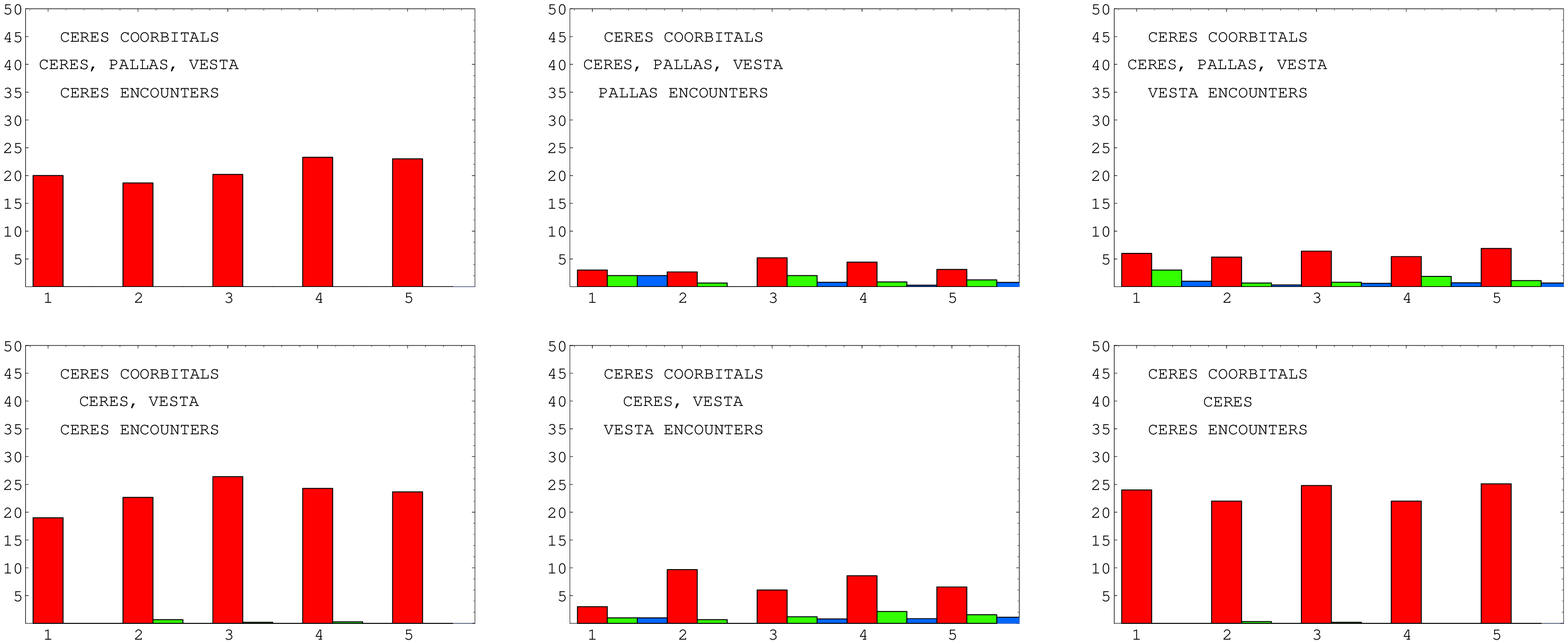}
\caption[Histograms of numbers of encounters of Ceres co-orbitals with massive asteroids as a function 
of encounter distance in the three models described in Section 5.]{Christou and Wiegert 2010, Coorbitals of Ceres and Vesta}
\label{fig:nencdistceres}
\end{figure}

\begin{figure}
\centering
\includegraphics[width=15cm,angle=0]{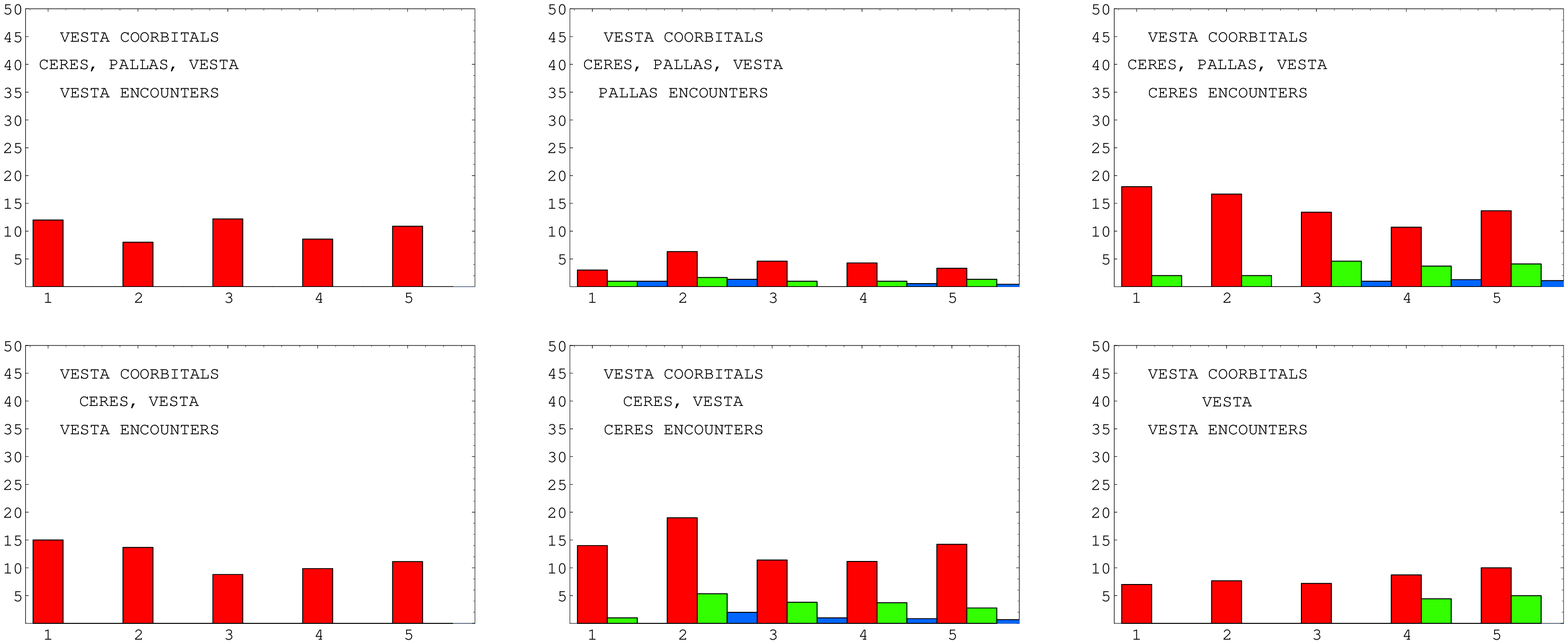}
\caption[As Fig.~\ref{fig:nencdistceres} but for Vesta co-orbitals.]{Christou and Wiegert 2010, Coorbitals of Ceres and Vesta}
\label{fig:nencdistvesta}
\end{figure}

\begin{figure}
\centering
\includegraphics[width=11cm,angle=0]{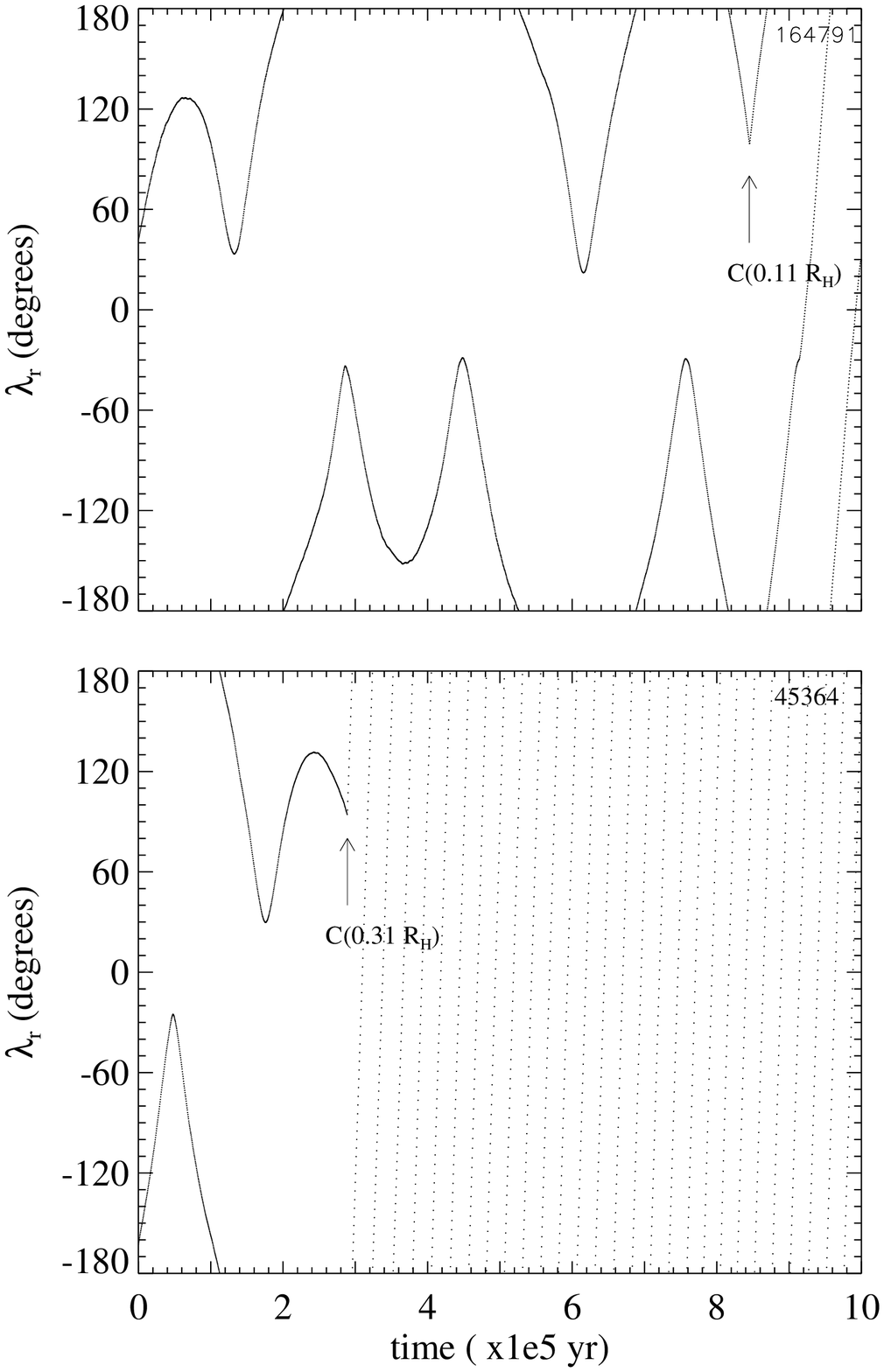}
\caption[Two examples of coorbitals of Vesta ejected from the resonance after close encounters with Ceres. 
The vertical arrows indicate the moment of encounter with Ceres (`C') while the number in brackets is 
the closest approach distance in $R_{\rm H}$.]{Christou and Wiegert 2010, Coorbitals of Ceres and Vesta}
\label{fig:encceres}
\end{figure}

\begin{figure}
\centering
\includegraphics[width=8cm,angle=0]{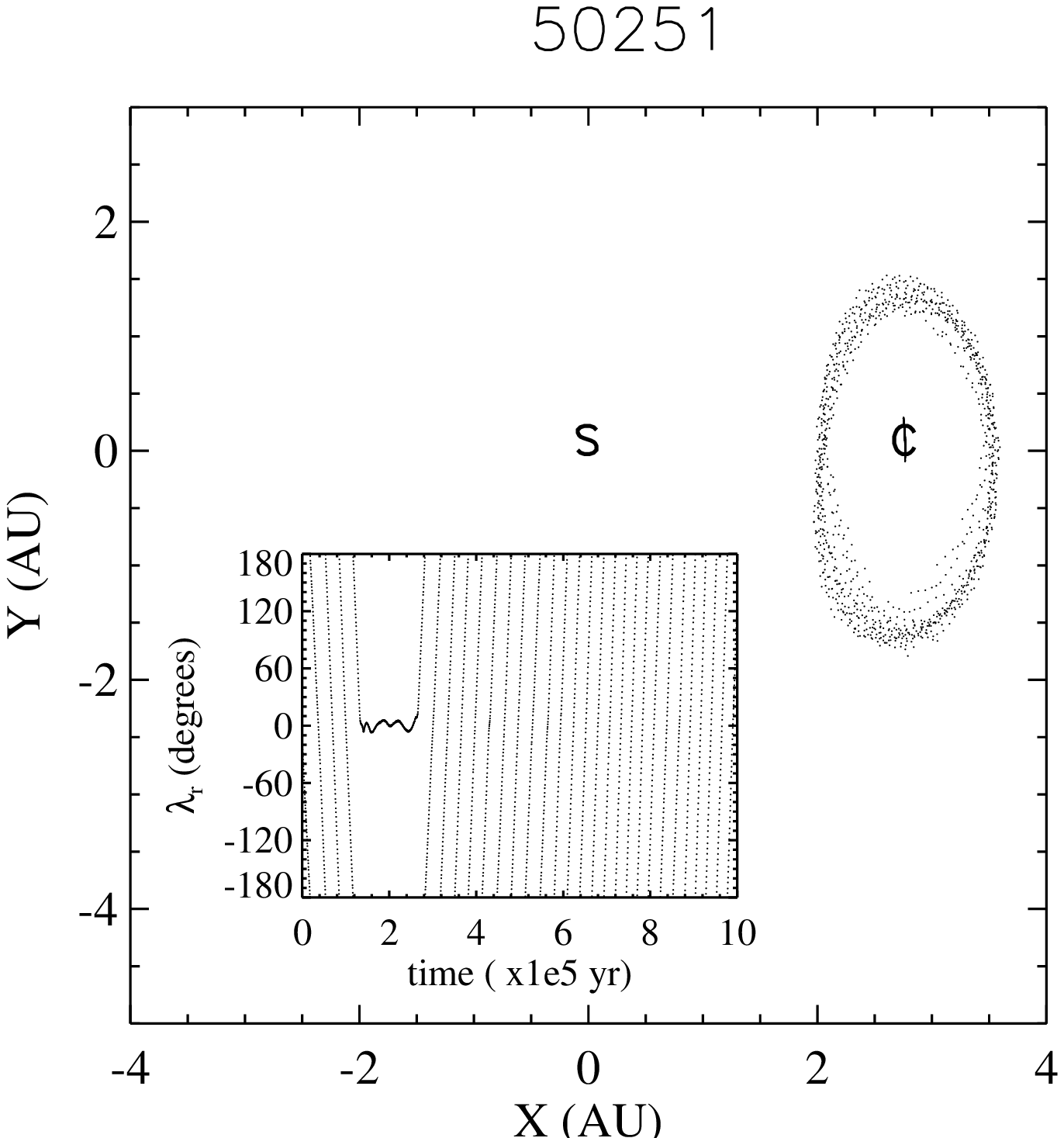}\includegraphics[width=8cm,angle=0]{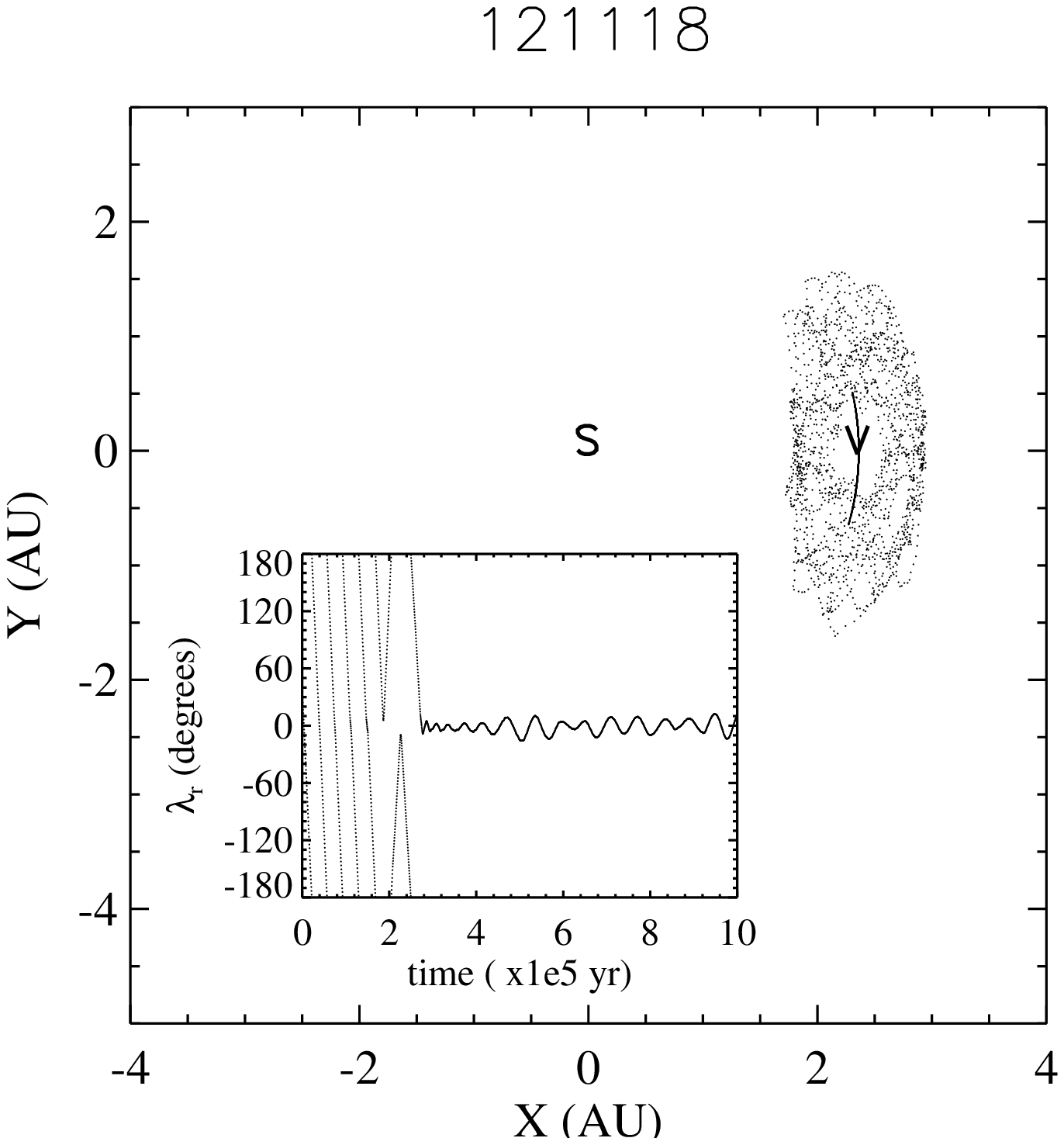}
\vspace{5cm}
\caption[Two examples of MBAs that exhibited Quasi-Satellite libration with Ceres and Vesta in our simulations. The position of the Sun is indicated by the character `S'.]{Christou and Wiegert 2010, Coorbitals of Ceres and Vesta}
\label{fig:quasisats}
\end{figure}

\begin{figure}
\centering
\includegraphics[width=8cm,angle=0]{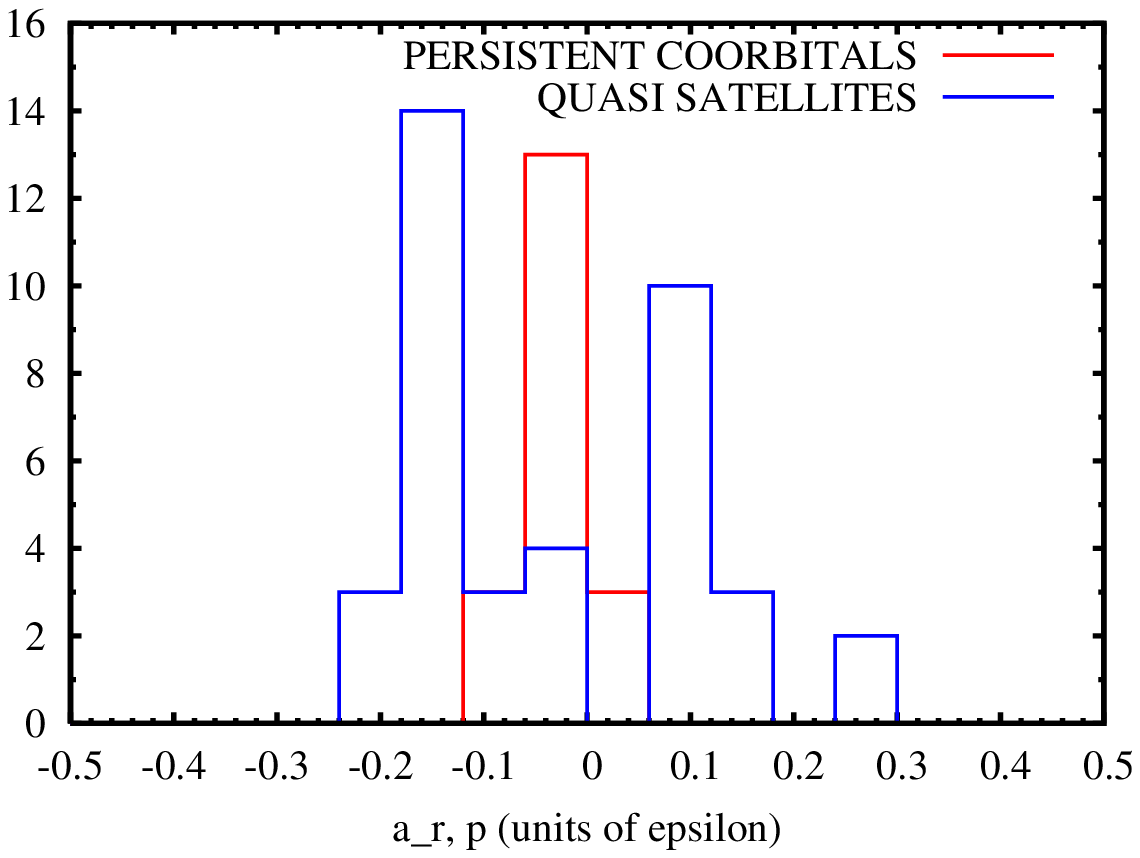}\includegraphics[width=8cm,angle=0]{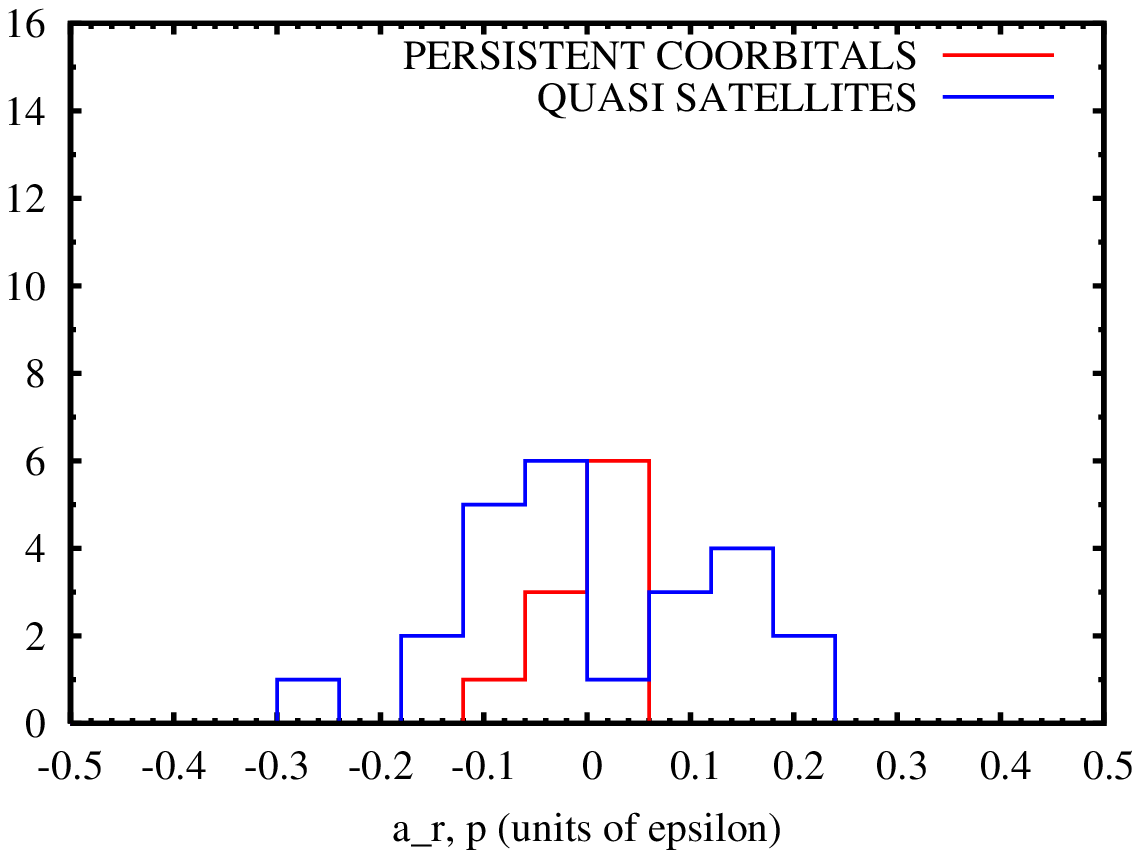}
\caption[Histogram of the relative proper semimajor axis (in units of $\epsilon$) of those MBAs 
that persisted in coorbital libration (blue line) and those that became temporary quasi-satellites 
(red line) of Ceres (left) and Vesta (right).]{Christou and Wiegert 2010, Coorbitals of Ceres and Vesta}
\label{fig:qsats_vs_a}
\end{figure}

Statistically, we find that 39 (24) out of the 129 (94) Ceres (Vesta) co-orbitals exhibited QS motion at some 
point during the $2 \times 10^{6}$ yr period covered by the simulations reported in Section 3, a fraction of 
25-30\% in both cases. In the case of Vesta we find three episodes of QS libration of unusually long 
($> 4 \times 10^{5}$ yr) duration. One is that of 121118 illustrated in Fig.~\ref{fig:quasisats}, 
the others concern MBAs 22668 and 134633. In Fig.~\ref{fig:qsats_vs_a} we compare the distribution of 
the relative proper semimajor axes of all objects that became temporary QS librators with the members 
of the persistent coorbital population. The two are generally separate 
with the former population further away and on either side of the latter. We believe this is because
MBAs capable of becoming QSs are highly energetic. In other words, and referring to the top left and top 
right panels of Fig.~\ref{fig:model_fit}, the value of the energy integral $3C / 8 \mu$ - represented by the horizontal
line - is generally well separated from the potential $S$. Hence the amplitude of 
$a_{r}$ libration is only small near $\lambda_{r} =0^{\circ}$ i.e. if the object is in QS mode at $t=0$. 
This is the case for the one current and the two imminent QS orbiters of Ceres: 76146, 138245 and 166529. 
The values of $a_{r}$ for these objects are $-$0.0068, $-$0.0155 and $-$0.0083 respectively.

A particularly interesting subtype of QS librator is that with vanishing guiding centre amplitude. 
These are the ``retrograde satellites'' (RS orbits) of \citet{Jackson1913}. In that case, the motion can be 
studied through an energy integral analogous to Eq.~\ref{zvc} but valid only in Hill's approximation to motion 
near the secondary. It depends explictly on $e_{r}$, $I_{r}$, $\omega_{r}$ and $\Omega_{r}$ (cf Eqs.~28 and 29 
of N99).

Activating the QS mode requires $e_{r} \gg \epsilon$ while physical proximity of the object to the secondary 
in an RS orbit is controlled by $e_{r}$, $I_{r}$ and $\omega_{r}$. For the small values of $\epsilon$ considered 
here, the relative eccentricity can still be small in absolute terms (eg 10$\epsilon$ or $\sim 10^{-2}$). If $I_{r}$ 
is also small \cite[typically $< e_{r}$ for bound orbits; see][]{Namouni1999}, then such objects can remain, in 
principle, within a few times $10^{6}$ km of Ceres or Vesta. Note that small $e_{r}$ implies a small libration 
amplitude for $\lambda_{r}$ since $\lambda_{r} < e_{r}$ (see Section 3.3 of N99). The long-term stability of these 
configurations depends, through the energy integral, on the evolution of the relative orbital elements. 
If these vary slowly, the asteroid remains trapped in QS motion for many libration cycles. 

To model the secular evolution of the asteroid's orbit we need a theoretical model of co-orbital motion 
within an N-body system. Such models do exist (eg the theory of \citealp{Message1966} valid near L4 and L5 and the more general model of \citealp{Morais1999,Morais2001}). However,
the case in hand violates several assumptions for which those theories are strictly valid: low to moderate $e$ and $I$ 
of the asteroid (here $e\mbox{, }I \gg \epsilon$), a coorbital mode (QS) that was not examined in those works and a 
clear separation of timescales between coorbital motion and the secular evolution of the orbit with 
$\dot{\varpi}\mbox{, }\dot{\Omega} \ll \dot{\lambda}$ (in fact, here we observe that 
$\dot{\lambda} < \dot{\varpi}\mbox{, }\dot{\Omega}$). To check that the secular evolution of
$e$, $I$, $\varpi$ and $\Omega$ of the asteroids does not depend on the presence of absence of the secondary mass
 we integrated the orbit of Vesta coorbital MBA 139168 with the eight major planets
both with and without the massive asteroids (including Vesta) for 1 Myr. We found that the differences in
the eccentricity and inclination vectors - $e \exp {\rm i} \varpi$ and $I \exp {\rm i} \Omega$ respectively - are $<5\%$
between the two cases. 
In addition, the asteroids' elements do not exhibit any of the features that arise from the three-body dynamics \citep{Namouni1999}.
Hence, it is reasonable to assume that the secular forcing of $e_{r}$ and $I_{r}$ is fully decoupled from the coorbital dynamics
and can be modelled by N-body secular theory.


In the absence of mean motion resonances, the secular evolution of the eccentricity and inclination vectors 
within a system of N bodies in near-circular, near-planar orbits around a central mass can be approximated 
by the so-called Laplace-Lagrange system of 2N first order coupled differential equations which are linear 
in $e$ and $I$ (Murray \& Dermott 1999). The corresponding secular solution for the eccentricity and inclination vectors of a particle introduced into that system has the form of a sum of N+1-periodic complex functions:
\begin{eqnarray}
\mathbf{e} &=& e_{f} \exp{\rm i} \left(g_{f} t + \beta_{f} \right) + \sum^{N}_{i=1} E_{i} \exp {\rm i} \left(g_{i} t + \beta_{i}\right)  \\
\mathbf{I} &=&  I_{f} \exp{\rm i} \left(s_{f} t + \gamma_{f} \right)+\sum^{N}_{i=1} I_{i} \exp {\rm i} \left(s_{i} t + \gamma_{i}\right) 
\end{eqnarray}
where the first and second terms on the right-hand-side are referred to as the {\it free} and {\it forced} 
components respectively. The parameters of the forced component are derived directly from the Laplace-Lagrange 
solution. They, as well as the free eigenfrequencies $g_{f}$ and $s_{f}$, are dependent only on the 
semimajor axes of the planets and the particle.  
 
 As with the theory of Morais, the validity of these expressions is limited to low-to-moderate $e$, $I$ and $g_{f} \neq g_{i}$, 
$s_{f} \neq s_{i}$. Brouwer \& van Woerkom (1950) showed that the 5:2 near-resonance between Jupiter 
and Saturn modifies the Laplace-Lagrange parameters of the real solar system.
However, the independence of the forced component on the particle's $e$ and $I$ still holds; this allows us to write
 \begin{eqnarray}
 \mathbf{e}_{r} &=&  e_{f} \exp{\rm i} \left(g_{f} t + \beta_{f} \right)  -  e^{\rm \star}_{f} \exp{\rm i} \left(g^{\rm \star}_{f} t + \beta^{\rm \star}_{f} \right) \\
 \mathbf{I}_{r} &=&  I_{f} \exp{\rm i} \left(s_{f} t + \gamma_{f} \right)  -  I^{\rm \star}_{f} \exp{\rm i} \left(s^{\rm \star}_{f} t + \gamma^{\rm \star}_{f} \right) 
 \end{eqnarray}
where the superscript `$\star$' refers to Ceres or Vesta as appropriate. Hence, $e_{r}$ and $I_{r}$ are given by
\begin{eqnarray}
 e_{r} &=&  {e}^{2}_{f}  +  {e^{\star}_{f}}^{2} + 2 e_{f}  e^{\rm \star}_{f} \cos \left[\left(g_{f} - g^{\rm \star}_{f}\right) t + \left( \beta_{f} - \beta^{\rm \star}_{f} \right) \right] \\
I_{r} &=&  I^{2}_{f} +  {I^{\star}_{f}}^{2} + 2 I_{f}  I^{\rm \star}_{f} \cos \left[\left(s_{f} - s^{\rm \star}_{f}\right) t + \left( \gamma_{f} - \gamma^{\rm \star}_{f} \right) \right]\cdot
 \end{eqnarray} 
The requirement for slow-evolving $e_{r}$ and $I_{r}$ implies $g_{f} \simeq g^{\rm \star}_{f}$, 
$s_{f} \simeq s^{\rm \star}_{f}$. Further, a necessary but not sufficient condition for $e_{r}$ and 
$I_{r}$ to be small is $e_{f} \simeq e^{\rm \star}_{f}$, $I_{f} \simeq I^{\rm \star}_{f}$. 
For these to be {\it concurrently} small, the following condition must also hold 
\begin{equation}
\frac{\pi - \Delta \beta_{f}}{\pi - \Delta \gamma_{f}}=\frac{\Delta g_{f}}{\Delta s_{f}}
\label{phases}
\end{equation}
 where $\Delta \beta_{f} = \beta_{f} - \beta^{\rm \star}_{f}$, $\Delta \gamma_{f} = \gamma_{f} - \gamma^{\rm \star}_{f}$, 
$\Delta g_{f} = g_{f} - g^{\rm \star}_{f}$ and $\Delta s_{f} = s_{f} - s^{\rm \star}_{f}$.
All these criteria, except the last one, can be tested for by making use of proper elements. The last criterion involves the proper phases which are not included 
in the database but can, in principle, be recovered through harmonic analysis of the orbital element time series.

As an example, we specify an upper limit in  $\Delta e_{f}$ and  $\Delta \sin I_{f}$ of 0.01 and a limit 
in $\Delta g_{f}$ and $\Delta s_{f}$ of 1 arscec per year. For Ceres co-orbitals, these correspond to 
a maximum epicycle excursion of $\sim 0.05$ AU and a period of $>1$ Myr in the evolution of $e_{r}$ and $I_{r}$. 
15 and 7 of the 39 QS librators identified in this work satisfy either of these criteria respectively but none 
do both. On the other hand, 2 coorbitals of Vesta do satisfy both criteria (3 and 8 respectively satisfy either one). Although neither of these two objects (78994 and 139168) satisfies the condition for concurrently small $e_{r}$ and $I_{r}$ we find the dynamical evolution of 139168 particularly interesting and we discuss it in some detail below.

\begin{figure}
\centering
\includegraphics[width=11cm,angle=0]{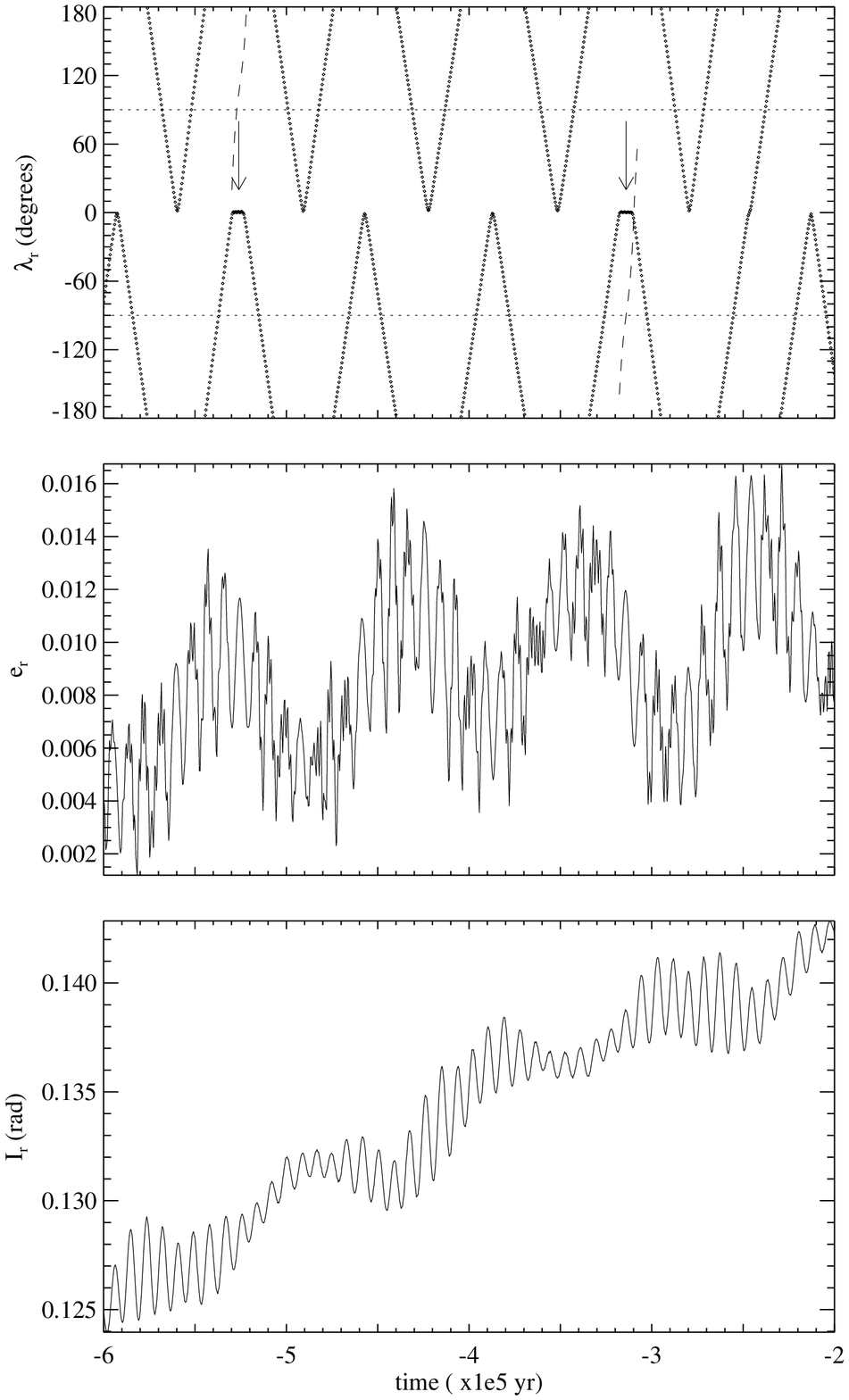}
\caption[Detail of the past dynamical evolution of 139168 in our simulations. 
The arrows indicate instances when the asteroid became an eccentric retrograde satellite of Vesta 
with $\lambda_{r}$ almost stationary at $0^{\circ}$.]{Christou and Wiegert 2010, Coorbitals of Ceres and Vesta}
\label{fig:139168_lei}
\end{figure}

\begin{figure}
\centering
\includegraphics[width=13cm,angle=0]{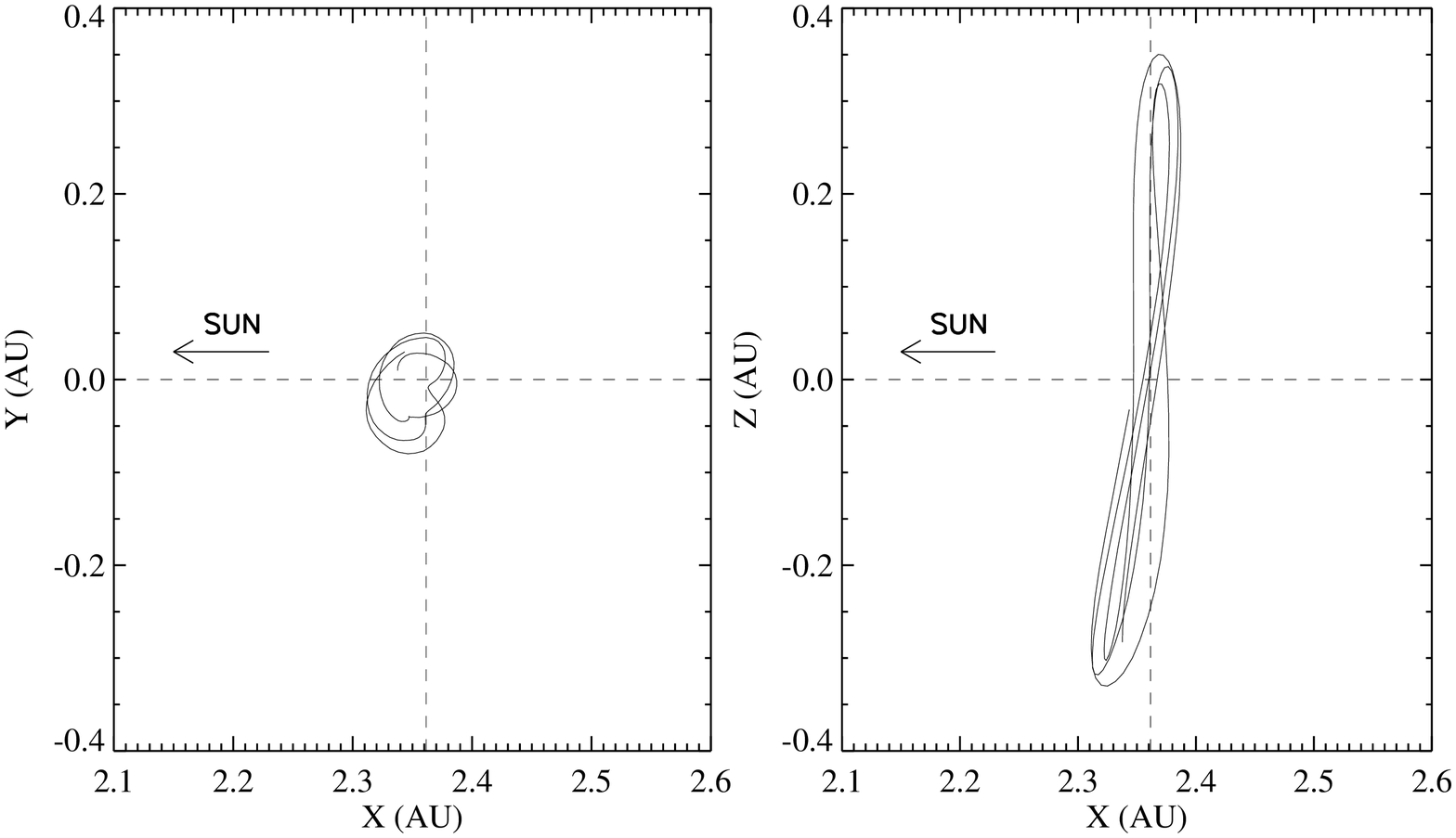}
\caption[Motion of 139168 with respect to Vesta in cartesian ecliptic coordinates during an episode 
of QS capture at $t \simeq - 3 \times 10^{5}$ yr indicated in Fig.~\ref{fig:139168_lei}. The dashed lines mark the location $x=a_{p\mbox{, }{\rm Vesta}}\simeq2.3615$ au, $y=z=0$.
Note the large amplitude of the vertical motion ( right panel) compared to the planar projection 
(left panel).]{Christou and Wiegert 2010, Coorbitals of Ceres and Vesta}
\label{fig:139168_xyz}
\end{figure}

This asteroid has $\Delta e_{f} = -0.0029379$, $\Delta \sin I_{f} = -0.0006901$, $\Delta g_{f}= 0.020303$ 
arcsec/yr and $\Delta s_{f} = 0.075063$ arcsec/yr. Inspection of our simulations of the nominal orbit of 
the asteroid shows that currently $e_{r} \sim 0.013$, $\sin I_{r} \sim 0.15$. Both elements are slowly 
increasing in time from the present with a period significantly longer than the $2 \times 10^{6}$ yr spanned 
by our integrations. Towards the past, The relative inclination continues to decrease with $\sin I_{r} \sim 0.10$
at $t=- 10^{6}$ yr while  $e_{r}$ reaches a minimum of $< 0.008$ at $t=-8.5 \times 10^{5}$ yr.
For most of the integrated timespan, the object is a horseshoe with a very small opening angle, suggesting 
that it is sufficiently energetic to enter a QS mode (top left panel of Fig.~\ref{fig:model_fit}) with $3 C / 8 \mu \sim 3.5$. 
Indeed, we find two instances, indicated by the vertical arrows in the top panel of Fig.~\ref{fig:139168_lei}, 
when the asteroid is trapped into a QS mode for $10^{4}$ yr and an $\lambda_{r}$ amplitude of 
$\sim 1^{\circ}$ ($\sim 10^{-2}$ rad). Since this is comparable to $e_{r}$ (middle panel), these are not, 
strictly speaking, RS orbits. Nevertheless, it implies that the planar component of the motion 
occurs within a few times $\lambda_{r} a$ of Vesta. In Fig.~\ref{fig:139168_xyz} we show the asteroid's 
motion for the phase of QS libration at $t=-3.15 \times 10^{5}$ yr in a heliocentric cartesian frame co-rotating 
with Vesta's mean motion. In the XY plane (bottom panel), the distance from Vesta varies between $10^{-2}$ and 
$8 \times 10^{-2}$ AU. However, the excursion in Z (bottom panel) is significantly larger, $\sim 0.3$ AU. 
Note that the centre of the planar motion is offset from Vesta's position. This is partly because one is 
looking at the superposition of two harmonic modes (guiding centre and epicycle) of comparable amplitude. 
In addition, the potential minimum may not be exactly at $\lambda_{r}=0$ since the maxima that bracket it 
are generally not equal unless $\omega_{r}=k \pi$. This becomes apparent if one expands
the Hill potential to order higher than 2 or utilises the full potential of the averaged motion i.e.~Eq.~\ref{zvc}.

Since $e_{r} < I_{r}$ the QS-enabling potential minimum at the origin exists only for values of $\omega_{r}$ 
sufficiently far from $k \pi$ while the maxima on either side are highest for $\omega_{r}=k \pi + \pi/2$, 
becoming singularities if $I_{r}=0$. These are the $K > 2$ orbits of N99. We can verify that this is the case 
here by overplotting $\omega_{r}$ on the top panel of Fig.~\ref{fig:139168_lei} (dashed curve). We find that, 
during QS libration, the value of $\omega_{r}$ is near $\pm 90^{\circ}$ (horizontal dotted lines) as expected.  

Eventually, $e_{r}$ and $I_{r}$ will pass through their minima sufficiently closely in time to make long-term 
capture in a low amplitude QS mode possible. For the given values of $\Delta g_{f}$ and $\Delta s_{f}$ 
this should occur every $\sim 6.4 \times 10^{7}$ yr. A crude estimate on the expected number of such objects 
may be made under the assumptions that (a) this is the only object of its type, (b) the proper element catalog 
is complete for MBAs with $a<2.8$ AU down to an absolute magnitude $H \simeq 16$ and (c) the duration of QS 
capture is $\geq 10^{5}$ yr, similar to large amplitude QS phases observed for other asteroids. The resulting 
average frequency of such objects at any one time is $\sim 1.5 \times 10^{-3}$. Adopting an absolute magnitude 
distribution law of $10^{0.3 H}$ \citep{Gladman.et.al2009}, we find that the frequency reaches unity for 
$H \sim 26$ or $D=12-37$ m objects. Unlike large amplitude Quasi-Satellites, the cartesian velocity of such 
objects with respect to Vesta is not high. For 139168 it ranges from 2.2 to 3.2 km $\mbox{\rm sec}^{-1}$ 
for the integration spanning the last 1 Myr and will be lower if $I_{r}$ is smaller. Hence, and in view of possible
{\it in situ} satellite searches by missions such as DAWN \citep{Cellino.et.al2006}, newly discovered 
objects in apparent proximity to Vesta in the sky would need to be carefully followed up to determine whether 
they are ``true'' (ie keplerian) satellites or small-amplitude Quasi-Satellites. 
 
\section{Conclusions and Discussion}
In this work we have demonstrated the existence of a population of Main Belt Asteroids (MBAs) in the coorbital 
resonance with the large asteroid (4) Vesta and the dwarf planet (1) Ceres. Libration within the resonance is 
transient in nature; our integrations show that these episodes can last for $>2 \times 10^6$ yr. Partly due to 
the significant eccentricities and inclinations of these asteroids, we find that their dynamics are similar 
to those that govern the evolution of near-Earth asteroids in the 1:1 resonance with the Earth and Venus. 
However, due to the high density of objects as a function of semimajor axis, a steady state population of 
$\sim 50$ co-orbitals of Ceres and $\sim 45$ of Vesta is maintained. Apart from the natural dynamics of eccentric 
and inclined 1:1 librators, we identify two other mechanisms which contribute to the temporary nature of these 
objects. One is the inherent chaoticity of the orbits; with some exceptions, particles started in neighbouring 
orbits evolve apart over $10^{5}$ yr timescales.
The other is close encounters with massive asteroids that do not participate in the Sun-Secondary-Particle three 
body problem. We show individual cases of asteroids leaving the coorbital resonance with Vesta following a deep 
encounter with Ceres. It is not clear if the latter effect is significant at the population level. It may be 
adversely affecting the occurence of long-lived tadpole librators of Vesta. Finally, we show that bound or 
Quasi-Satellite orbits around both Ceres and Vesta can exist and identify 3 current Quasi-Satellite librators 
around Ceres as well as an object which may experience episodes of long-lived bound motion within a few times 
$10^{-2}$ AU from Vesta.

The demonstration of a resonant mechanism within the asteroid belt which acts independently of the major planets 
raises some interesting questions to be addressed by future work. One of these is whether there exists a 
threshold below which a mass becomes a particle in real planetary systems. In partial response to this 
question - and as part of the integrations reported in Section 3 - we simulated the motion of several MBAs with 
proper semimajor axes within the coorbital region of 2 Pallas and 10 Hygiea. None of these was trapped in 
co-orbital libration which leads us to conclude that this threshold for the solar system's Main Asteroid Belt 
is $\sim 10^{-10}$ solar masses. However, this may not be true of other planetary systems with fewer 
and/or less massive planets. We speculate that co-orbital trapping of planetesimals by terrestrial planets or 
large asteroids may give rise to observationally verifiable dynamical structures. This is not a new idea 
\citep[eg][]{Moldovan.et.al2010} but our results show that even small bodies ($10^{-10}<\mu<10^{-6}$) can 
maintain transient populations of Trojans in a steady state and that the dynamical excitation of planetesimals 
in a disk does not necessarily imply that the coorbital resonance becomes ineffective. In this context, it may be 
relevant to the dynamical evolution of the larger planetesimals in protoplanetary disks and the planets or protoplanet 
cores embedded within them \citep{Papaloizou.et.al2007}.    

\section*{Acknowledgements}
The authors would like to thank Dr Fathi Namouni for kindly answering
our numerous questions regarding eccentric and inclined coorbital motion. 
Part of this work was carried out during a visit of AAC at UWO, 
funded by NASA's Meteoroid Environment Office (MEO). Astronomical research at the Armagh Observatory 
is funded by the Northern Ireland Department of Culture, Arts and Leisure 
(DCAL).
\newpage


\bibliographystyle{elsarticle-harv}
\bibliography{I11975Final}
\clearpage
\protect
\listoffigures

\renewcommand{\baselinestretch}{1.0}
\protect
\newpage
\begin{table}[htb]
\centering
\caption[Dynamical parameters of large asteroids considered in this work. See text for details.]{Dynamical parameters of large asteroids considered in this work. See text for details.}
\begin{tabular}{lcccc}
\noalign{\smallskip}
\hline \hline
\noalign{\smallskip}
        &     Ceres   &  Vesta         &    Pallas        &    Hygiea  \\ \hline
$\mu \mbox{ }\times 10^{10}$   &  4.699         &   1.358           &     1.026            &     0.454     \\   
$\epsilon  \mbox{ }\times 10^{4} $  &    5.42         &    3.56            &       3.25           &   2.47         \\
$a_{\rm proper}$ (AU)   &   2.7670962          &      2.3615126          &      2.7709176            &   3.1417827         \\ 
$\epsilon a_{\rm proper}  \left(\times 10^{4}  \mbox{AU}\right)$ &  15.0  &  8.4    &   9.0             &    7.8       \\ 
\noalign{\smallskip} \hline \hline
\end{tabular} 
\label{tab:massive}
\end{table} 
\newpage
\begin{table}[htb]
\centering
\caption[Statistical results of the numerical simulations reported in Section 3]{
Statistical results of the numerical simulations reported in Section 3. See text for details.}
\begin{tabular}{lcccc}
\noalign{\smallskip}
\hline \hline
\noalign{\smallskip}
         &    Full       &     Now        &     $\mbox{L}_{4}$ Tadpole  &   $\mbox{L}_{5}$ Tadpole  \\\hline
Ceres    &   129         &     51         &      15 (2)     &       4 (2)     \\
Vesta    &    94         &     44         &       5         &       7 (1)     \\\hline
         &   Horseshoe   &   Transition   &    QS           &     Always    \\\hline
Ceres    &    20         &     11         &       1         &      18        \\
Vesta    &    21         &     11         &       0         &      10        \\
\noalign{\smallskip} \hline \hline
\end{tabular} 
\label{tab:statistics}
\end{table} 
\newpage
\begin{table}[htb]
\centering
\caption[Estimated mean and Full Width at Half Maximum (FWHM) of the distributions presented in Fig.~\ref{fig:histo_ap2}.]{Estimated mean and Full Width at Half Maximum (FWHM) of distributions presented in Fig.~\ref{fig:histo_ap2}.}
\begin{tabular}{lcccccc}
\noalign{\smallskip}
\hline \hline
\noalign{\smallskip}
         &  \multicolumn{2}{c}{All} &  \multicolumn{2}{c}{Current} & \multicolumn{2}{c}{Persistent} \\
         &  \multicolumn{2}{c}{ coorbitals} &  \multicolumn{2}{c}{coorbitals} & \multicolumn{2}{c}{coorbitals} \\
         &     $\mu$     &   FWHM              &  $\mu$          &                FWHM   &  $\mu$     &                FWHM     \\\hline \noalign{\smallskip}
Ceres    &  $ {-0.029}^{ \pm 0.011}$  & ${0.341}^{ \pm 0.021} $ & ${-0.019}^{ \pm 0.002}$  & ${0.169}^{ \pm 0.005 }$  &  ${-0.023}^{ \pm 0.001}$ & ${0.096}^{ \pm 0.001} $\\
Vesta    &  $ {-0.006}^{ \pm 0.009}$  & ${0.301}^{ \pm 0.019}$  & $ {-0.011}^{ \pm 0.005}$ & ${0.151}^{ \pm 0.009}$   &  ${-0.010}^{ \pm 0.001}$ & ${0.052}^{ \pm 0.002}$\\
\noalign{\smallskip} \hline \hline
\end{tabular} 
\label{tab:mufwhm}
\end{table} 
\newpage
\begin{table}[htb]
\centering
\caption[Statistical results of the numerical simulations for the models described in Section 5. See text for details.]{Results of the numerical simulations for the models described in Section 5. See text for details.}
\begin{tabular}{lccccccc}
\noalign{\smallskip}
\hline \hline
\noalign{\smallskip}
       \multicolumn{2}{c}{Secondary}         &  \multicolumn{2}{c}{} &  \multicolumn{2}{c}{} &  \multicolumn{2}{c}{All Massive }  \\
       \multicolumn{2}{c}{Body}              &  \multicolumn{2}{c}{Ceres/Vesta Only} &  \multicolumn{2}{c}{No Pallas} &  \multicolumn{2}{c}{Asteroids}  \\\noalign{\smallskip}
         \multicolumn{2}{c}{}                &    Persistent  &  Single Mode   &   Persistent  & Single Mode    &  Persistent    & Single Mode\\\hline\noalign{\smallskip}
        Ceres               &                &      26/129    &     6/129      &      28/129   &   14/129       &      32/129    &   15/129\\
        Vesta               &                &      23/94     &     11/94      &     18/94     &    3/94        &      18/94     &   6/94  \\
\noalign{\smallskip} \hline \hline
\end{tabular} 
\label{tab:stability}
\end{table} 
\newpage

\renewcommand{\baselinestretch}{2.0}

\newpage
\clearpage
\clearpage
\clearpage
\clearpage
\clearpage
\clearpage
\clearpage
\clearpage
\clearpage
\clearpage
\clearpage
\clearpage
\clearpage
\clearpage
\clearpage
\clearpage
\clearpage
\clearpage
\clearpage
\end{document}